\documentclass[aps,prb,superscriptaddress,twocolumn,10pt,longbibliography]{revtex4-2}

\usepackage{amsmath}
\usepackage{graphicx}
\usepackage{bm}
\usepackage{color}

\date{\today}

\definecolor{darkblue}{rgb}{0.1,0.2,0.6}
\definecolor{darkred}{rgb}{0.8,0.1,0.2}
\usepackage[colorlinks,citecolor=darkblue,linkcolor=darkred,urlcolor=darkblue]{hyperref}
\usepackage[all]{hypcap}

\usepackage{etoolbox}
\usepackage{mathtools}
\usepackage{amsmath}
\usepackage{amsfonts}
\usepackage{amssymb}
\usepackage[normalem]{ulem}

\begin{document}

\title{Evolution of many-body systems under ancilla quantum measurements}
\author{Elmer~V.~H.~Doggen}
\email[Corresponding author: ]{elmer.doggen@kit.edu}
\affiliation{\mbox{Institute for Quantum Materials and Technologies, Karlsruhe Institute of Technology, 76021 Karlsruhe, Germany}}
\affiliation{\mbox{Institut f\"ur Theorie der Kondensierten Materie, Karlsruhe Institute of Technology, 76128 Karlsruhe, Germany}}
\author{Yuval Gefen}
\affiliation{Department of Condensed Matter Physics, Weizmann Institute of Science, 7610001 Rehovot, Israel}
\author{Igor V.~Gornyi}
\affiliation{\mbox{Institute for Quantum Materials and Technologies, Karlsruhe Institute of Technology, 76021 Karlsruhe, Germany}}
\affiliation{\mbox{Institut f\"ur Theorie der Kondensierten Materie, Karlsruhe Institute of Technology, 76128 Karlsruhe, Germany}}
\author{Alexander D.~Mirlin}
\affiliation{\mbox{Institute for Quantum Materials and Technologies, Karlsruhe Institute of Technology, 76021 Karlsruhe, Germany}}
\affiliation{\mbox{Institut f\"ur Theorie der Kondensierten Materie, Karlsruhe Institute of Technology, 76128 Karlsruhe, Germany}}
\author{Dmitry G.~Polyakov}
\affiliation{\mbox{Institute for Quantum Materials and Technologies, Karlsruhe Institute of Technology, 76021 Karlsruhe, Germany}}

\begin{abstract}
Measurement-induced phase transitions are the subject of intense current research, both from an experimental and a theoretical perspective. We explore the concept of implementing quantum measurements by coupling a many-body lattice system to an ancillary degree of freedom (implemented using two additional sites), on which projective measurements are performed.
We analyze the effect of repeated (``stroboscopic'') measurements on the dynamical correlations of interacting hard-core bosons in a one-dimensional chain. An important distinctive ingredient of the protocol is the fact that the detector ancillas are not re-initialized after each measurement step. The detector thus maintains memory of the accumulated influence by the measured correlated system. Initially, we consider a model in which the ancilla is coupled to a single lattice site. This setup allows obtaining information about the system through Rabi oscillations in the ancillary degrees of freedom, modulated by the ancilla-system interaction. The statistics of quantum trajectories exhibits a ``quantum-Zeno-valve effect'' that occurs when the measurement becomes strong, with sharp branching between low and high entanglement. 
We proceed by extending numerical simulations to the case of two ancillas and, then, to measurements on all sites. With this realistic measurement apparatus, we find evidence of a disentangling-entangling measurement-induced transition as was previously observed in more abstract models. The dynamics features a broad distribution of the entanglement entropy.
\end{abstract}

\maketitle

\section{Introduction}

The concept of measurements in quantum systems has been a topic of great interest ever since the dawn of quantum mechanics, but the microscopic description of the dynamics of systems under the influence of measurements has remained a challenging problem \cite{Zurek2003a, Schlosshauer2005a, Wiseman2009, Jacobs2014}. In recent years, this problem has gained much attention particularly because of its relevance to the stability of quantum computing architectures \cite{Aharonov2000a,Preskill2018a,Bharti2022} and to the goal of achieving ``quantum supremacy'' \cite{Arute2019a, Noh2020a, Zhou2020a, Wu21, Zhong21, Pan22, Wellnitz2022a}. Furthermore, the physics of quantum measurements is closely connected with the study of open quantum systems \cite{Breuer2002, Rotter2015a}. It is also fundamentally relevant to understanding the link between the microscopic physics and the thermodynamics on macroscopic scales \cite{Polkovnikov2011a}.

Recently, the notion of a measurement-induced entanglement transition---a dynamical phase transition driven by the strength or frequency of measurements---has been proposed \cite{Skinner2019a,Li2018a}. Initially discussed in the context of quantum circuits \cite{Skinner2019a,Li2018a,Chan2019a, Szyniszewski2019a, Li2019a,  Bao2020a, Choi2020a, Gullans2020a, Gullans2020b, Jian2020a, Zabalo2020a, Iaconis2020a, Turkeshi2020a, Zhang2020c, Nahum2021a, Ippoliti2021a, Ippoliti2021b, Lavasani2021a, Lavasani2021b, Sang2021a, Fisher2022,  Kelly2022a, Liu2022a, Weinstein2022a, Sierant2023a, Jian2023}, the idea of measurement-induced transitions has been extended to non-interacting lattice fermions \cite{Cao2019a, Alberton2020a, Chen2020a, Tang2021a, Coppola2022, Ladewig2022, Carollo2022, Buchhold2022, Yang2022, Marcin2022, Fava2023}, Dirac fermions \cite{Buchhold2021a}, the quantum Ising model \cite{Lang2020a, Rossini2020a, Biella2021a, Turkeshi2021, Tirrito2022, Yang2023, Weinstein2023, Murciano2023}, the Sachdev-Ye-Kitaev model \cite{Jian2021a}, as well as a number of other integrable \cite{Minoguchi2022} and nonintegrable many-body systems \cite{Tang2020a, Fuji2020a, Goto2020a, Lunt2020a, Doggen2022a, VanRegemortel2021a, Altland2022,Block2022a, Sierant2022a, Minato2022a, Yamamoto2023, Silveri2023}. Evidence of a measurement-induced transition has also been reported in trapped-ion \cite{Noel2022a} and superconducting \cite{Koh2022} quantum-processor architectures. 

An intriguing question is to what extent the essential properties of these measurement-induced transitions are universal \cite{Vasseur2019a, Szyniszewski2020a, Bao2020a, Lu2021a, Gopalakrishnan2021a, Fava2023}, and whether a sharp transition exists at all for specific models \cite{Cao2019a, Bao2020a, Coppola2022, Carollo2022}. 
In particular, it has been debated whether the main features of the transitions are sensitive to the system realization (e.g., quantum circuits vs. Hamiltonians with or without interactions) and to implementation of measurements or monitoring. This includes a distinction between the protocols comprising rare strong measurements and those with frequent (or continuous) weak measurements. These different classes of protocols may possess identical long-time dynamics at the level of averaged density matrix, yet this level is inappropriate for studying entanglement dynamics.
It is worth noting that, in the vast majority of works in the field, either projective local measurements are employed, or an effective model of a continuous monitoring of the system's dynamics is introduced (e.g., a stochastic Schr\"odinger equation formulated in terms of the system's degrees of freedom).       

A realistic description of the measurement process \cite{Wiseman2009,Jacobs2014} requires a microscopic consideration of the joint evolution of the measured system and the detectors. In this situation, without resetting (re-initializing) the detectors, not only the backaction of measurement (present in all types of measurements) affects the systems, but also an inevitable accumulated feedback of the system on the detector can bias the next measurements. This may influence the classification of generalized measurements into strong- or weak-measurement classes, as the effective measurement strength depends on the system's state along the quantum trajectory. This type of ``memory-effect'' correlations is absent in conventional models implementing measurements or monitoring. The correlated dynamics in the physical realization of the coupled system-detector setup can therefore be expected to exhibit novel features. 

A possible interplay of various types of criticality and correlations in measured systems \cite{Ashida2016, Lunt2020a,  Garratt2022a, Yang2023, Weinstein2023, Murciano2023, Sun2023, Yamamoto2023} is another important facet of the problem.
In this context, one may draw a certain analogy between a measurement-induced entanglement transition for a finite density of detectors and disorder-induced localization transition. At the same time, it is known that even a single impurity drastically affects the properties of one-dimensional correlated systems ~\cite{KaneFisherPRB92, GiamarchiBook}. For example, a weak impurity may cut a wire into two, suppressing transport, similar to localization. This calls for the consideration of the effect of a single detector on the correlated chain, especially when the detector (or a pair of detectors) is located near the bipartition cut that is used for the definition of entanglement.
In particular, it is tempting to look for some features characteristic of entanglement transitions in models with a minimum number of detectors, which would shed more light on the nature of the true transition.
Furthermore, one may expect that, in a realistic measurement setup, the ancilla-system coupling could be ``renormalized'' by the correlation in the main chain, similar to the Kane-Fisher impurity problem \cite{KaneFisherPRB92}, which in turn would affect the entanglement transition. 
The above analogies with correlated wires with impurities serve as an additional motivation for our study. 

In this work, we investigate the influence of measurements on quantum many-body lattice systems by coupling the system to ancillary degrees of freedom. The projective measurements are performed on the ancillary sites only. A distinguishing feature of our approach is that, after projecting the ancillary sites, the latter are not re-initialized for the next measurement cycle. The effect of the measurements on the dynamics of the ``main" system is mediated by interactions between it and the ancilla (``detector''), with no particle exchange between the two. Within the proposed framework, the dynamics of the system as a whole (main system plus ancilla) remains  closed, aside from the projections of the detector sites at fixed discrete intervals. The projections are effected through nonunitary operators that implement Born's rule, which is the only assumption we make about the nature of the measurement process. Furthermore, the feedback effect of the main system on the dynamics of ancillary degrees of freedom is automatically taken into account in the course of joint evolution between consecutive projections. 
This procedure thus brings us a step closer to a realistic description of quantum measurements in interacting many-body systems, with the ancillary sites mimicking a ``measurement apparatus.'' As such, our protocol is applicable to an arbitrary many-body lattice system and can readily be adapted to current experimental settings of interest, including cold atoms, trapped ions, and superconducting qubits.

We first demonstrate the power of this approach by outlining how the ``measured'' site density of interacting hard-core bosons on a one-dimensional lattice can be reconstructed from the ancilla dynamics. 
Next, we investigate the effect of a measurement backaction on the density distribution and the entanglement entropy, for one and two sites coupled to the detectors. The feedback of the main system on the detectors gives rise to what we dub a ``quantum-Zeno-valve effect'' (involving a blockade of the ancilla dynamics), which occurs either when the stroboscopic-projection frequency is commensurate with the ancilla Rabi frequency or at strong system-detector coupling.   
Finally, we analyze the dynamics of the system for a finite density of measured sites. We observe manifestations of a disentangling-entangling measurement-induced transition in the time- and system size-dependence of the entanglement entropy averaged over realizations of individual quantum trajectories. 
The dynamics features a broad distribution of the
entanglement entropy and density fluctuations. We thus provide evidence for a measurement-induced transition driven by the change of ancilla-system coupling in a correlated quantum system.

The paper is organized as follows. We formulate the model and specify the measurement protocol in Sec.~\ref{s2}. In Sec.~\ref{sec:rabi}, we outline the basic concept of ``measuring'' the main system by means of performing projections on the ancilla. In Sec.~\ref{sec:repmeas}, we study the effect of backaction from repeated measurements with one or two ancillas, focusing on the density distribution and the entanglement entropy. Two different initial states of the main system are considered: a domain-wall state in Sec.~\ref{sec:singlepair} and the ground state in Sec.~\ref{sec:dmrg_gs}. 
In Sec.~\ref{sec:5D}, we present results for the entanglement and density dynamics in a chain with all sites coupled to the detectors, with a particular focus on the disentangling-entangling transition.
Section \ref{s6} provides a summary and conclusions. Some of the technical details of performed numerical simulations and additional benchmarks are described in Appendices.

\section{Model, measurement protocol, and observables}
\label{s2}

\subsection{Model}

For concreteness, we consider a model of hard-core bosons on a one-dimensional lattice at half filling. The Hamiltonian of the chain (main system) is given by
\begin{equation}
\mathcal{H}_\mathrm{s} =  \sum_{i=1}^{L-1} \left[ -\frac{J}{2} \left(b_{i}^\dagger b_{i+1} + \mathrm{H.c.}\right)  + U\hat{n}_{i} \hat{n}_{i+1} \right],
\label{eq:ham}
\end{equation}
where $b_i$ ($b_i^\dagger$) annihilates (creates) a boson on site $i$, the lattice size is $L$, $\hat{n}_i \equiv b_i^\dagger b_i$ denotes the density operator on site $i$, $J$ is the strength of hopping between neighboring sites, and $U$ is the strength of interaction between particles on neighboring sites. This model is equivalent to the XXZ spin chain, and its ground-state and dynamical properties have been studied extensively \cite{Znidaric2011a, Laflorencie2016a}.

\begin{figure}
\centering
\includegraphics[width=\columnwidth]{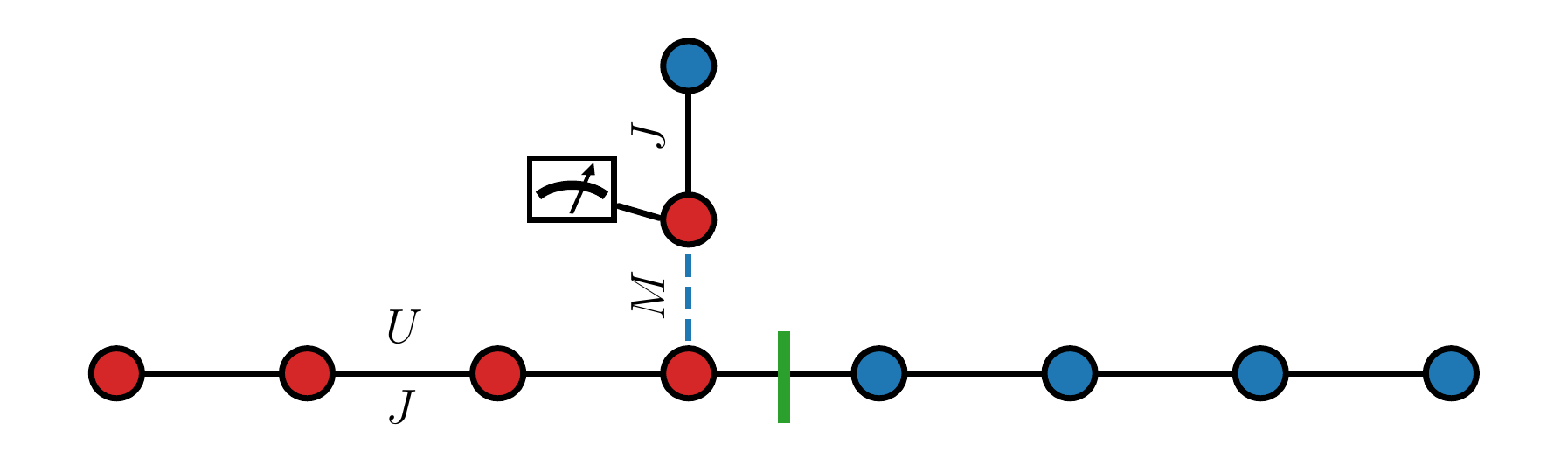}
\caption{Schematic depiction of the model and a ``domain-wall'' initial condition for the case of a single ancilla (implemented as a pair of the ancillary sites shown vertically). Hard-core bosons can hop across the main chain (horizontal row, depicted with $L=8$ sites for the purpose of illustration) with the hopping amplitude $J$, in the presence of nearest-neighbor interactions of strength $U$. The chain is coupled to the ancilla pair of sites through the interaction $M$. No particle transport between the ancilla and the main chain is permitted. The occupation on the lower site of the ancilla pair is projectively measured (as indicated by the dial screen) at discrete intervals. The ancilla pair accommodates a single boson that can hop between its two sites with the same hopping amplitude $J$ as for the bosons in the main chain. Initially, the sites colored in red and blue are, respectively, fully occupied and empty. The ancilla state is not reinitialized following each projective measurement, but rather follows unitary evolution interspersed with the projections. 
Specifically, after the projection, the ancilla is set to either the $|01\rangle$ or $|10\rangle$ state, depending on the outcome of a Born-rule measurement. The green line indicates bipartition for the entanglement entropy (both ancilla sites belong to the left part of the entire system).}
\label{fig:diag}
\end{figure}

The simplest setup with an ancilla is then realized when the ancilla is represented by a single pair of sites. Taking $L$ to be even, we couple, by means of interaction, site $L/2$ of the main system and one of the sites of an ancillary pair, as illustrated in Fig.~\ref{fig:diag}. The total Hamiltonian $\mathcal{H}$ is then written as $\mathcal{H} = \mathcal{H}_\mathrm{s} + \mathcal{H}_\mathrm{a}+\mathcal{H}_\mathrm{sa}$, where
\begin{equation}
\mathcal{H}_\mathrm{a} =  -\frac{J}{2}\left( a_1^\dagger a_2 + a_2^\dagger a_1 \right) 
\label{eq:meas_ham}
\end{equation}
describes the ancilla and
\begin{equation}
\mathcal{H}_\mathrm{sa} = M \hat{n}_{L/2} a^\dagger_1 a_1  
\label{eq:meas_ham_sa}
\end{equation}
the coupling between the main system and ancilla. Here, $a_{1,2}$ ($a^\dagger_{1,2}$) are the annihilation (creation) operators of hard-core bosons on ancilla sites 1 and 2, and $M$ is the strength of interaction between the main system and the ancilla. 
Note that the choice of interaction in Eqs.~(\ref{eq:ham}) and (\ref{eq:meas_ham_sa}) in the form of only particle-particle interaction breaks particle-hole symmetry of the model. Everywhere below, the intra-chain and chain-ancilla interaction is taken to be repulsive ($U>0$ and $M>0$). For simplicity we have also chosen, in Eq.~(\ref{eq:meas_ham}), the hopping strength in the ancilla to be the same as in the main system. 

\begin{figure}
\centering
\includegraphics[width=\columnwidth]{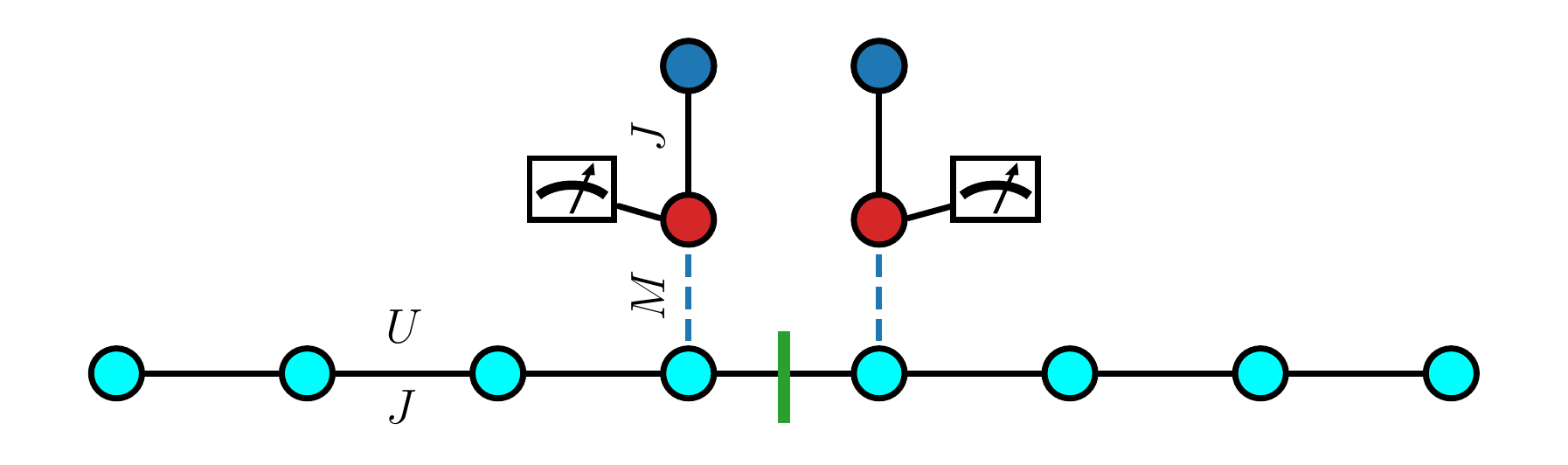}
\caption{Setup and initial state with two ancilla pairs. The notation is the same as in Fig.~\ref{fig:diag}. The main chain, for the purpose of illustration shown for $L = 8$, is initialized in the ground state for $M=0$ (with density $n=1/2$, cyan circles) and coupled to two ancilla pairs at two sites in the middle. Each of the ancilla pairs is initialized in a state with the site that is connected to the main chain being occupied. The green line denotes entanglement bipartition cut.}
\label{fig:diag_dmrg}
\end{figure}

In what follows, we first study the basic setup with a single ancillary pair, as described above. We proceed by including a second ancillary pair, as shown in Fig.~\ref{fig:diag_dmrg}, permitting a more detailed understanding of the influence of measurements on the system.
Finally, we employ this approach to the case where the ancillary sites are attached to every site of the chain, see Fig.~\ref{fig:projdiag_multi}. 

\subsection{Ancilla measurement protocol}
\label{ss3}

We perform projective density measurements on the lower ancilla site (the one connected to the main chain through interaction), periodically at an interval $\Delta T$. Consider the case of a single ancilla (multiple ancillas are treated similarly). After each projection, the ancilla pair is in a state with either the lower site occupied or the lower site empty, i.e., the whole system is then in the state  $|\Psi \rangle_{10} \otimes |1 0 \rangle$ or $|\Psi \rangle_{01} \otimes |0 1\rangle$, respectively, where $|10\rangle$ and $|01\rangle$ denote the ancilla states, and $|\Psi \rangle_{10}$ and $|\Psi \rangle_{01}$ are the corresponding states of the main chain. Because of the backaction of the measurement onto the main chain, the states $|\Psi \rangle_{10}$ and $|\Psi \rangle_{01}$ are in general different. The probability of each outcome is determined by the Born rule according to the density $n_a = \langle a_1^\dagger a_1 \rangle$. Note that the projection breaks both unitarity and integrability.

\begin{figure}
    \centering
    \includegraphics[width=\columnwidth]{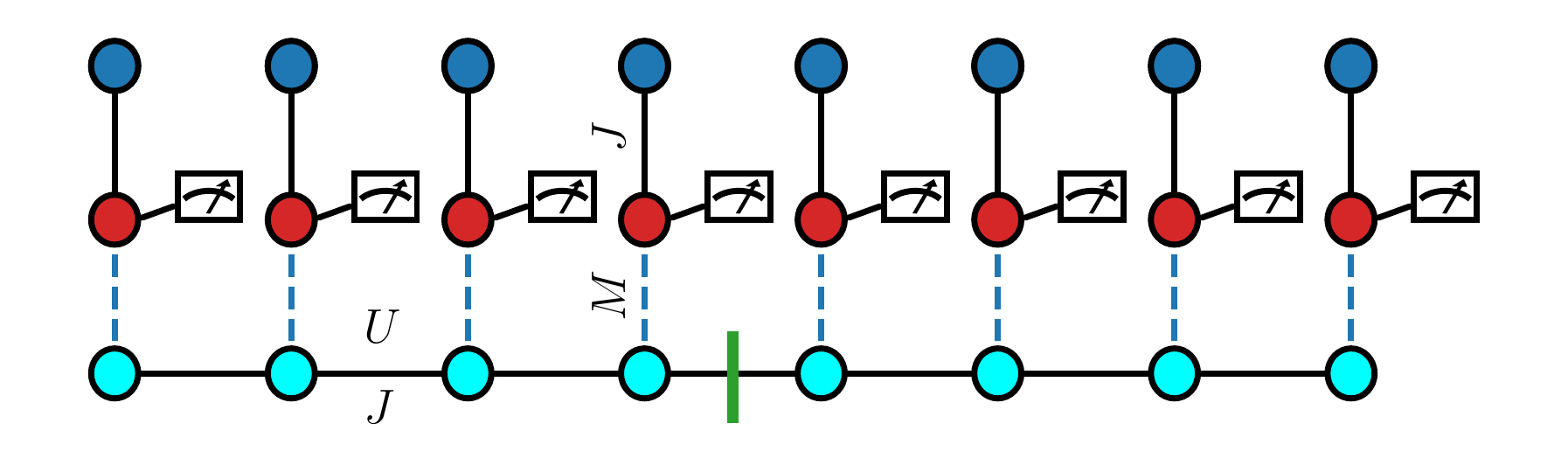}
    \caption{Setup and initial state with $L$ ancilla pairs. The main chain (cyan circles), for the purpose of illustration shown for \mbox{$L = 8$}, is initialized in the  ground state for $M=0$ and coupled to an ancilla pair at every site. Each of the ancilla pairs is initialized in a state with the site that is connected to the main chain being occupied. The green line denotes entanglement bipartition.}
    \label{fig:projdiag_multi}
\end{figure}

We model the projection as an instantaneous event. Importantly, the probability of a particular readout shown by the ``measurement apparatus'' (the ancilla pair) is dependent on the whole history of the measurement process. Indeed, because of the Born rule, the previous measurement results affect the dynamics in the main chain (backaction). This, in turn, affects the value of $n_a$ at the time when the projection takes place (feedback), and so on. This backaction-feedback memory loop gives rise to system-ancilla correlations. In particular, a self-sustained ``blockade'' or ``freezing" of the ancilla dynamics may occur, resulting in an almost projective measurement of the system site, which leads to a quantum Zeno-like effect. Numerically, the unitary time evolution of the entire system---main chain plus ancilla(s)---is obtained using the time-dependent variational principle \cite{Haegeman2016a, Paeckel2019a}, see Appendix \ref{sec:appendix_num}.
For the model with ancillas attached to every site of the chain
(Fig.~\ref{fig:projdiag_multi}), we move beyond exact simulations and utilize the full power of matrix-product-state (MPS) simulations, considering systems of up to $72$ sites, including ancillary sites.

It is worth mentioning that the implementation of generalized measurements through coupling the system to ancillary degrees of freedom followed by their projections has deep roots in various contexts involving weak measurements (see, e.g., Refs.~\cite{Gurvitz1997,Korotkov1999, Korotkov2001, Goan2001, Allahverdyan2004a,  Shpitalnik2008, Romito2008, Taranko2012, Sinitsyn2014, Barbarino2019, Bao2020a, Esin2020, Kumar2020, Roy2020, Herasymenko2021, Dhar2022}). 
In particular, this type of measurement allows one to associate the counting statistics \cite{Sinitsyn2014, Turkeshi2021, Dhar2022, Tirrito2022} of discrete outcomes of ancilla projections with the properties of the system without strong backaction. Furthermore, these outcomes can be used in a feedback loop to control the system by active-decision making on further measurements or the unitary evolution (see, e.g., Refs.~\cite{Herasymenko2021,Friedman2022,ODea2022a,Silveri2023} and references therein).  
Microscopically, perhaps, the closest setup resembling ours was introduced in Refs.~\cite{Gurvitz1997, Korotkov1999}, where a double-dot (``two-site'') system was electrostatically coupled to a point contact in a conducting channel (see also Refs.~\cite{Goan2001, Korotkov2001, Romito2008, Taranko2012} for related setups). However, the roles of ancilla and system were interchanged in that setup compared to ours (cf. Ref.~\cite{Barbarino2019}): the transmission of the conducting channel was measured strongly, yielding the information about the occupation of the dot connected to it. Let us also emphasize that recent experimental works on measurement-induced entanglement transitions \cite{Noel2022a,Koh2022} employ ancillas for performing measurements on the system of qubits.

In a recent work \cite{Bao2020a}, a setup with ancillas was considered theoretically, in the context of measurement-induced transitions, for a quantum circuit. There are, however, essential differences between our framework and the one studied in Ref.~\cite{Bao2020a}. A clear difference is that we deal with a Hamiltonian system undergoing real time evolution, and which respects particle number conservation. Another difference is in the dependence of the effective measurement strength on the history of the system, which is an inherent property of our model. By contrast, a newly initialized (in a prescribed state) ancilla is introduced in Ref.~\cite{Bao2020a} at every ``time step'' (at every layer of the quantum circuit), similarly to the steering protocols of Refs.~\cite{Roy2020,Herasymenko2021}. 
This is an important difference, particularly because adaptive feedback and ``preselection'' mechanism may prove essential, as argued in Refs.~\cite{Buchhold2022, Lu2022, Friedman2022a, Iadecola2022, Friedman2022, ODea2022a, Ravindranath2022a, Silveri2023}, for observing a measurement-induced phase transition. In our framework, the feedback emerges ``automatically'', through the mechanism of joint unitary evolution fixed by the Hamiltonian, with no re-initialization of the ancilla after the projections. This results in effective adaptive dynamics of the entire system. A further key difference with previous works is that we explore the density and entanglement dynamics in various setups, with one, two, and many detectors. 

\subsection{Observables}
\label{s5a}

We focus on the following observables: (i) the density distribution $n_i(t) = \langle b_i^\dagger b_i \rangle (t)$ of bosons along the main chain, as a function of time $t$, (ii) the density $n_a(t) = \langle a_1^\dagger a_1 \rangle (t)$ of the ancilla boson at the lower site (the one coupled to the main chain through the interaction with magnitude $M$), and (iii) the bipartite von Neumann entropy of entanglement $S(t)$. The latter is defined as
\begin{equation}
S(t)  = -\mathrm{Tr}\Big( \rho_A \ln \rho_A \Big), \quad \rho_A = \mathrm{Tr}_\mathrm{B} \rho ,
\end{equation}
where $\rho$ is the density matrix of the whole system, $\rho_A$ is the reduced density matrix ($\mathrm{Tr}_\mathrm{B}$ denotes a trace over subsystem $B$), and we place the bipartition between subsegments $A$ and $B$ of the system in the middle of the main chain. Note that for an ancilla, both its sites also belong to one of the subsystems $A$ or $B$. Since the ancilla is entangled with the main chain during the unitary evolution between the projections, the ancillary degrees of freedom contribute to the value of $S$. 

From the density in the main chain $n_i$, we can obtain the particle imbalance:
\begin{equation}
    \mathcal{I} = -L/4 + \sum_{i=1}^{L/2} \langle \hat{n}_i \rangle. 
    \label{eq:imba}
\end{equation}
This quantity measures the bipartite fluctuations with respect to a homogeneous distribution of particles in the main chain, so that $\mathcal{I} = 0$ corresponds to a homogeneous state, and $\mathcal{I} = \pm L/4$ is a state where all particles are in the left ($+$) or right ($-$) side of the chain respectively. Recall that since we consider the main chain at half filling, there are $L/2$ bosons present there.

All three quantities (i)-(iii) are, in principle, directly measurable in experiment; the imbalance was considered for instance in Ref.~\cite{Choi2016a} (defined in a slightly different way). The entanglement can also be measured in state-of-the-art experimental settings \cite{Islam2015, Lukin2019, Brydges2019a, Lewis-Swan19, Foss-Feig2022a}. Within the framework outlined in Sec.~\ref{sec:rabi}, a measurement apparatus in the form of an ancilla pair can be efficiently used to ``measure'' the density distribution by means of projections performed only on the ancilla.  

\section{Rabi oscillations in the ancilla coupled to the main chain}
\label{sec:rabi}

Before moving on to study the effect of ancilla projections on the main system, let us outline the basic concept of gaining information about the main system from the dynamics of the ancilla. It is based on the use of the fact that the probability of projection to either ``particle" or ``hole" in the ancilla dynamics is controlled by the expectation value of the ancilla density $n_a$, introduced in Sec.~\ref{s2}, right before the moment of projection. Since the ancilla site at which the projection is made is coupled by particle-particle interaction to the chain, the unitary dynamics of $n_a$ between consecutive projections provides information about the occupation of the chain site to which it is connected. In the following Sec.~\ref{sec:repmeas}, we analyze the effect of the ancilla on the main chain, as opposed to the effect of the main chain on the ancilla as we consider in this section.

In the absence of coupling between the ancilla and the main chain ($M=0$), the ancilla dynamics is that of a two-level system subjected to periodic projections. Specifically, the ancilla displays Rabi oscillations, with the frequency given by $|J|$, and is reset after each projection performed on the lower ancilla site in either the $|1 0\rangle$ or $|0 1\rangle$ state, after which the Rabi oscillations continue with the same frequency.

For $M\neq 0$, the interaction between the main chain and the ancilla modifies both the characteristic Rabi frequency and the oscillation amplitude. As an instructive simple example, consider the static mean-field approximation for the configuration of Fig.~\ref{fig:diag}. In this approximation, the density operator $\hat{n}_{L/2}$ for site $L/2$ in the Hamiltonian is replaced by its expectation value $n = \langle \hat{n}_{L/2} \rangle$ and its time dependence is neglected. The ancilla boson then ``feels'' the time-independent potential $Mn$. 
The Rabi frequency $\Omega$ for the oscillations of $n_a(t)$ becomes
\begin{equation}
\Omega = \sqrt{J^2+M^2n^2},
\label{eq:rabi}
\end{equation}
growing as $M$ is increased, while the amplitude $A$ of the oscillations is reduced:
\begin{equation}
A =  \frac{1}{2} \frac{J^2}{J^2+M^2n^2}.
\label{eq:rabiamp}
\end{equation}
Note that Eqs.~(\ref{eq:rabi}) and (\ref{eq:rabiamp}) correctly reproduce the limit of ${M=0}$, giving, in particular, ${A=1/2}$ for the unperturbed two-level system. Henceforth we set $\hbar = 1$, express energies in units where ${J=1}$, and time in units of $1/J$.

\begin{figure}
\centering
\includegraphics[width=\columnwidth]{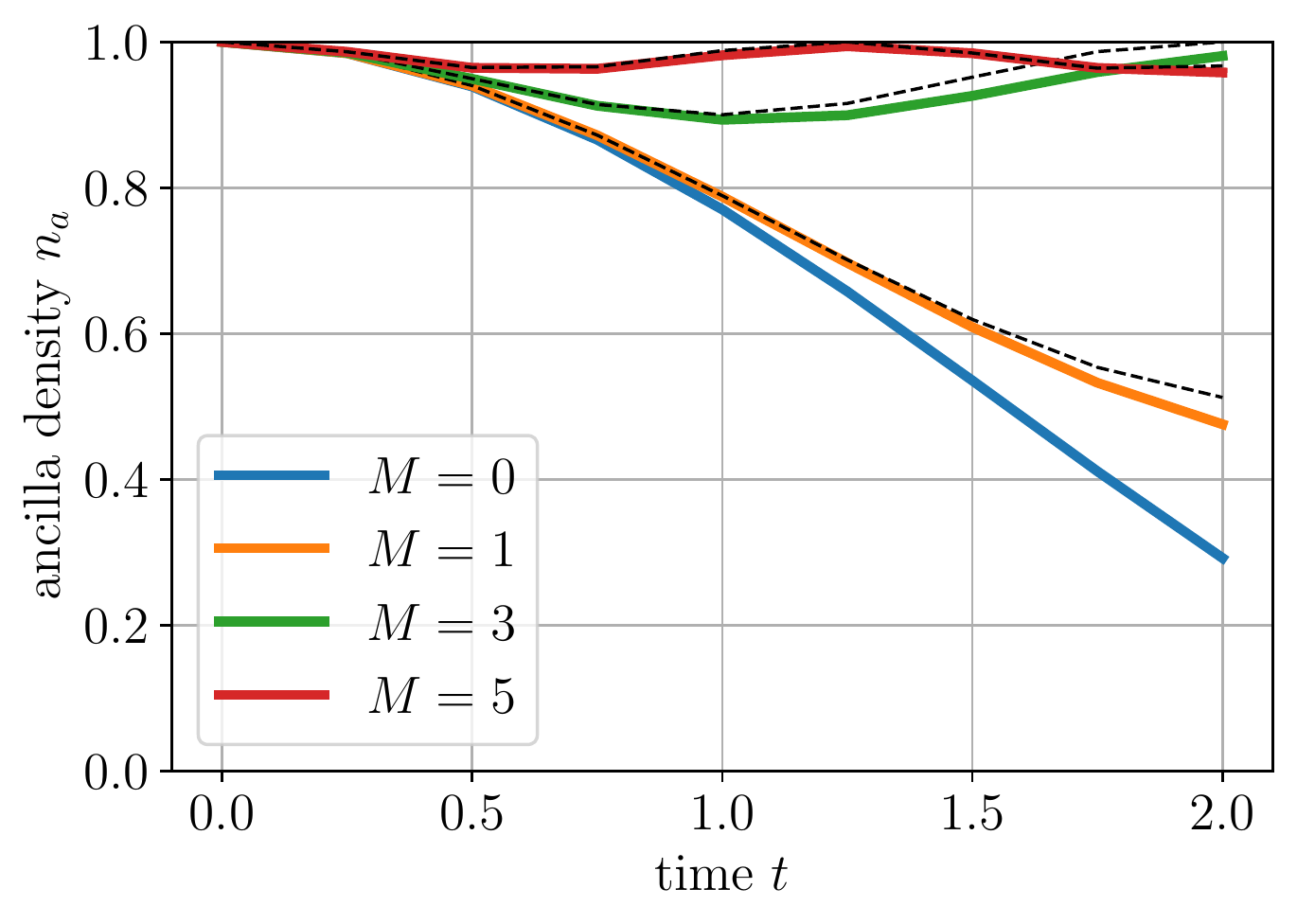}
\caption{Density $n_a$ of the lower ancilla site (connected to the main chain) as a function of time in the setup depicted in Fig.~\ref{fig:diag} (the ancilla is initially coupled to the occupied site of the chain, $n=1$), showing Rabi oscillations in the time interval before the first projection measurement ($\Delta T = 2$), starting with $n_a=1$, for $L = 16$ and $U = 1$. For $M=0$, the exact solution is shown; for $M \neq 0$, the result is obtained numerically. The dashed lines depict the two-level approximation (see the main text).}
\label{fig:Rabi}
\end{figure}

The ancilla dynamics in the setup of Fig.~\ref{fig:diag} (with the domain-wall initial state of the main chain) for $0<t<\Delta T$ and various choices of $M$ is shown in Fig.~\ref{fig:Rabi}, with the colored lines denoting the numerical results obtained from a full simulation of the entire system (for the case of $U=1$) and the dashed lines corresponding to Eqs.~\eqref{eq:rabi} and \eqref{eq:rabiamp}, for the case of site $L/2$ fully occupied ($n=1$) at $t=0$. The case of $M=5$ demonstrates a very good agreement between the numerical data and the two-level mean-field approximation. This is because the ancilla site in the limit of $M\gg 1$ acts as a pinning ``impurity" for the main chain. The small-amplitude Rabi oscillations in the ancilla are then a weak fast-oscillating perturbation of the otherwise slow dynamics of the entire system (tunneling processes in the main chain that change the occupation of the pinning site are weak for large $M$, yielding no appreciable change of $n$ for $M=5$ within the time interval $\Delta T$ in Fig.~\ref{fig:Rabi}). For intermediate $M$ (exemplified by the cases of $M=1$ and $M=3$), a noticeable deviation becomes apparent, resulting, apart from fluctuations around the mean-field solution, from the time evolution of $n$ associated with a melting of the domain wall. Qualitatively, however, Eqs.~(\ref{eq:rabi}) and (\ref{eq:rabiamp}), which become exact in the limits of both small and large $M$, capture well the leading time-dependent behavior in $n_a(t)$ for arbitrary $M$.

It is worth noting that the expectation value of the ancilla density for $M=3$ in Fig.~\ref{fig:Rabi} appears to be very close to 1 at the projection time $t=\Delta T =2$, similarly to $n_a$ for $M=5$, but in a sharp contrast to the case of $M=1$. This is because of an approximate resonant condition: $\Omega \Delta T=2\sqrt{10} \simeq 2\pi$ for $M=3$ and $n=1$. As a result, at the time of projection, the ancilla with an intermediate-strength coupling ($M=3$) is felt by the system as if the coupling would be extremely strong. In other words, the resonance condition makes the potential of the ancilla exerted on the bosons in the chain similar to the potential of a strong impurity. 
The probability of having no potential for a quantum trajectory ($a_1^\dagger a_1$ projected to 0) at $t=\Delta T$ vanishes upon approaching the exact resonance. This resonance is, however, in general broadened through the breakdown of the two-level approximation for the Rabi oscillations. We have checked that a similar behavior occurs also at other values of $M$ and $\Delta T$ satisfying approximately the resonant condition, in particular for $M=6$ and $\Delta T = 1$.

The dynamics in the chain breaks down the resonance condition by changing $n$ in Eq.~\eqref{eq:rabi}. 
However, as discussed above, a strong impurity (either with $M\gg 1$ or in the resonant case) slows down this dynamics, resulting in repeated measurements of $n=1$ (i.e., the quantum Zeno effect). Only rarely the occupations may change for the resonant coupling, which immediately removes the pinning, thus opening the channel underneath the ancilla. 
This behavior is reminiscent of the operation of a valve, hence we dub it the ``quantum-Zeno-valve effect" (see the detailed discussions of implications of this phenomenon in Sec.~\ref{sec:singlepair} below).

The procedure involves an inherent trade-off. On the one hand, the longer $\Delta T$, the stronger the effect of the interaction between the main system and the ancilla (the ``measurement apparatus''). On the other hand, as $\Delta T$ increases, the effect of the main chain on the ancilla is averaged over a longer time interval. This means that if one endeavors to obtain information about the main chain through subsequent measurements of the ancilla, this information will then be ``blurred'' over the time window $\Delta T$. In Fig.~\ref{fig:Rabi}, this effect is seen through the increasing difference in time between the simulated curves and those for the static mean-field approximation of Eqs.~(\ref{eq:rabi}) and (\ref{eq:rabiamp}). 

Another effect of a similar kind concerns the strength of interaction $M$. The characteristic Rabi frequency increases with growing $M$, i.e., the effect of interaction on the ancilla dynamics becomes apparent at earlier times for larger $M$ (Fig.~\ref{fig:Rabi}). At the same time, the effect of the ancilla on the main chain also increases as $M$ grows. In the limit of large $M$, the measurement of a particle at the ancilla site leads, as already mentioned above, to a complete freezing of dynamics in the main chain. Importantly, the backaction, like the projective measurement, is instantaneous and results in a ``spooky action at a distance'' at every site of the chain. In particular, the (expectation value of the) particle density at each site of the main chain changes discontinuously at each measurement step.

Building on the preceding discussion, the remarkably simple setup with a two-level ancilla can be utilized to ``measure'' $n$ in the main system through a subsequent determination of the ancilla dynamics. The interested reader is referred to Appendix \ref{sec:appendix_meas} for a proof-of-principle discussion concerning how such a measurement protocol might be effected in practice. In the following sections, we will focus on the effect of the ancilla projections on the main chain dynamics, and investigate possible signatures of a measurement-induced transition.

\section{Effect of repeated, spatially localized measurements on the main chain}
\label{sec:repmeas}

Having discussed the basic concept of coupling an ancilla to a many-body Hamiltonian system in Sec.~\ref{sec:rabi}, we now turn to an analysis of the effect of projective ancilla measurements on the main chain, where said measurements are performed repeatedly with a given frequency. We initialize the detectors by placing a boson in the ancillary pair at the lower site, with the other site empty. For the main chain, we consider two different initial states, both with half filling ($L/2$ bosons). We first examine, in Sec.~\ref{sec:singlepair}, the initial state in the form of a product state with a domain wall in the middle, such that the leftmost sites are fully occupied, as depicted in Fig.~\ref{fig:diag}. 
This initial state is especially convenient for exploring the effect of measurements on the relaxation of inhomogeneous density distribution, as characterized, in particular, by the particle imbalance, Eq.~(\ref{eq:imba}).
For this initial state, the measurement is performed with a single ancillary pair. 

We then proceed, in Sec.~\ref{sec:dmrg_gs}, to analyze the case where the initial state is the ground state of Eq.~\eqref{eq:ham} (with decoupled ancilla sites, $M=0$), and use a single ancilla, as well as two ancillary pairs at the neighboring sites of the main chain, as illustrated in Fig.~\ref{fig:diag_dmrg}.
Choosing this initial state has the benefit that the system, in the absence of measurements, does not evolve and remains in its weakly entangled state, as opposed to the domain-wall initial state. This is particularly useful for studying the entanglement dynamics.
The setting with two ancillas illuminates the nature of the feedback between the detectors in setups with multiple ancillas.  
Finally, in Sec.~\ref{sec:5D}, we will attach the ancillary pairs to every site of the main chain, which will be again initialized in its ground state.

\begin{figure*}
\centering
\includegraphics[width=.765\textwidth]{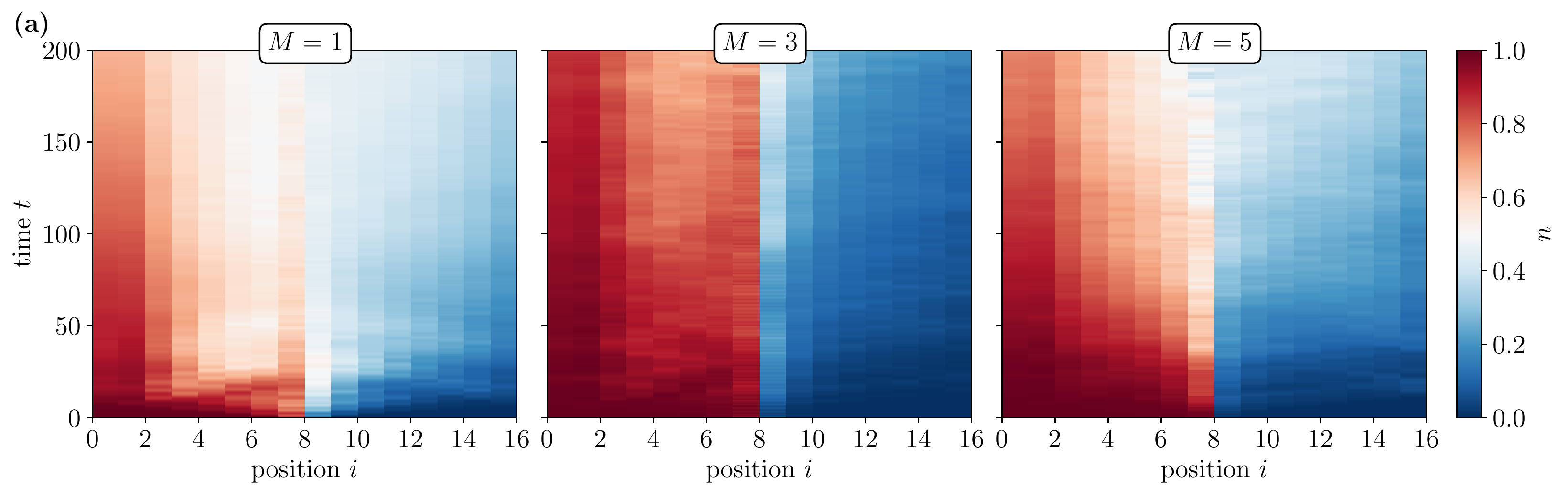}
\includegraphics[width=.225\textwidth]{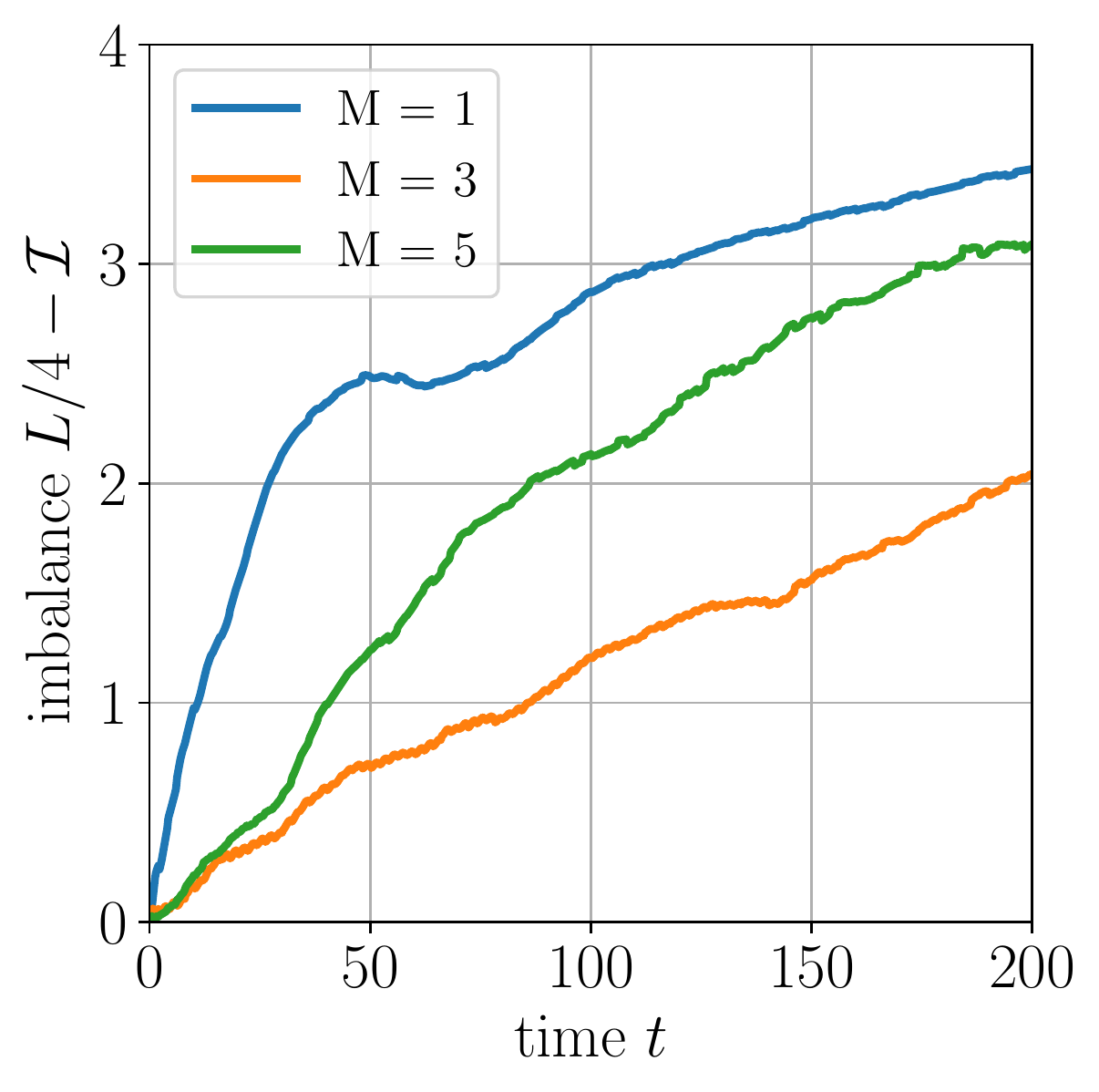}
\includegraphics[width=.275\textwidth]{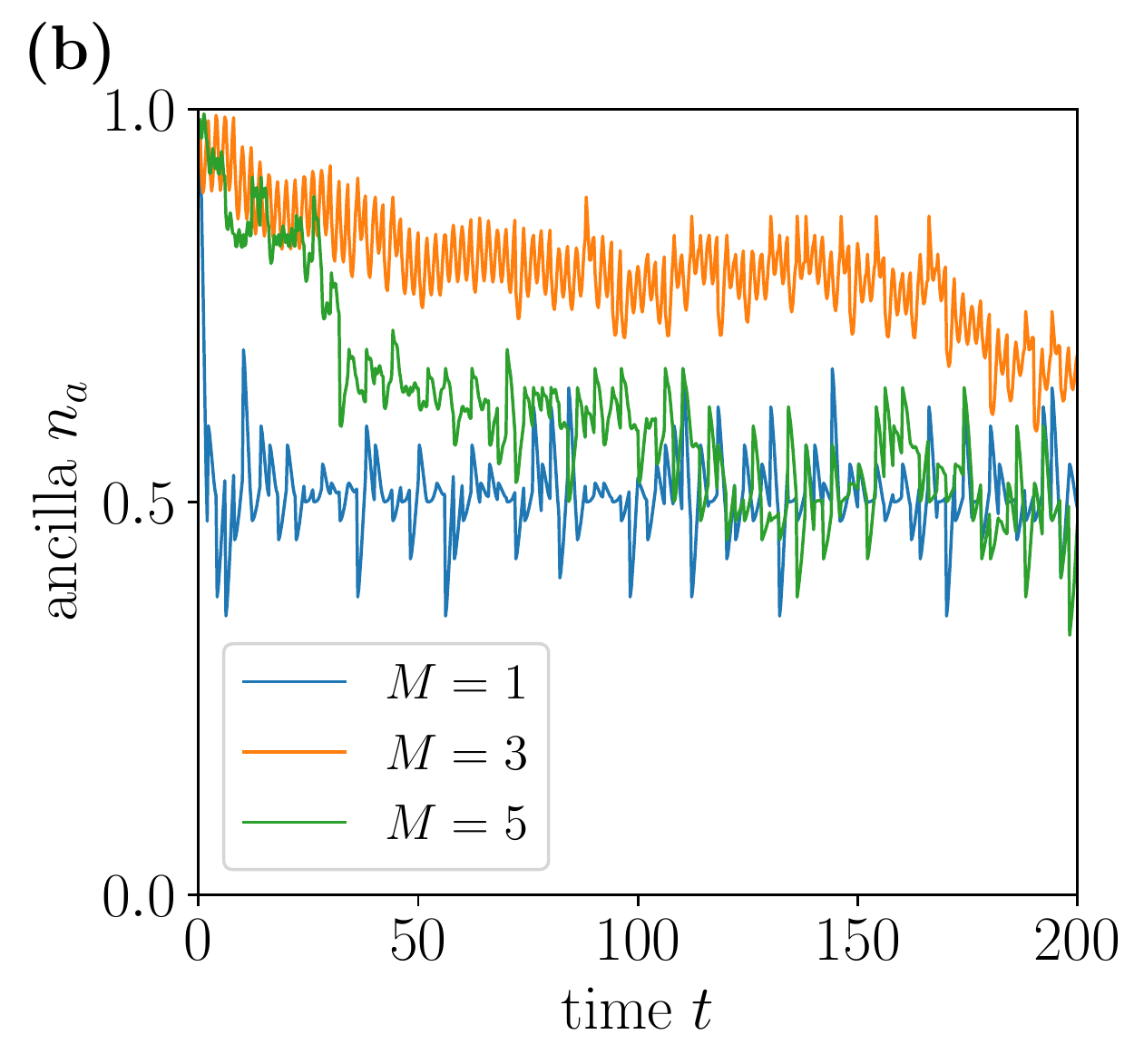}
\includegraphics[width=.715\textwidth]{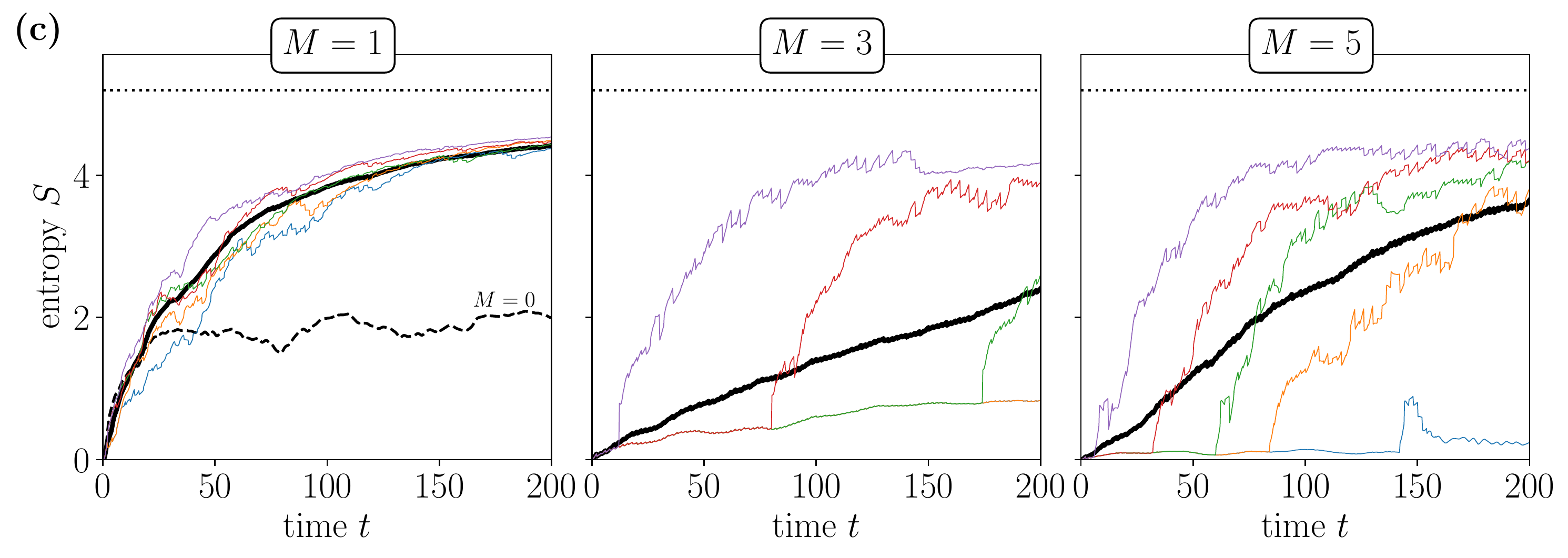}
\caption{Dynamics of the system schematically depicted in Fig.~\ref{fig:diag} (single ancilla; domain-wall initial state) for $L = 16$, $U=1$, and $M=1,3,5$. Measurement period $\Delta T = 2$. \textbf{(a)}: Particle density $n_i(t)$ in the main chain as a function of time $t$. Here and in other density plots below, the density $n_i$ (with $i=1\ldots 16$) is displayed in the interval $[i-1,i]$. The right panel shows the total number of particles that have moved from the left to the right half of the chain, as obtained through the density imbalance $\mathcal{I}$. \textbf{(b)}: Particle density $n_a$ at the lower ancilla site as a function of time, averaged over 40 realizations of quantum trajectories. \textbf{(c)}: von Neumann entropy of entanglement $S$ as a function of time. The black thick solid line depicts the average over realizations, similarly to (b). The five thin solid lines show dynamics of $S$ for individual quantum trajectories sorted by $\Xi = \int \mathrm{d}t \, S(t)$ [Eq.~\eqref{eq:entang_sorting}], with the highest value of $\Xi$ (purple curve), the lowest value (blue), and the values of $\Xi$ higher than those for 25 (orange), 50 (green), and 75 (red) percent of trajectories. The horizontal dotted line denotes $S$ for a thermal state in the thermodynamic limit (the Page entropy). The dashed line in the $M = 1$ panel shows the case of no coupling to the ancilla, $M = 0$, cf.~Fig.~\ref{fig:app_nomeas}(a) in Appendix \ref{sec:appendix_unc}. The case of very strong interaction $M = 50$ is discussed in Appendix \ref{sec:appendix_dwall}, see Fig.~\ref{fig:domainwall_M50}. The entropy in this case remains so small that it would be indistinguishable from zero on the scale of panel (c).}
\label{fig:ms8L16}
\end{figure*}

\subsection{Domain-wall initial state with a single ancilla} 
\label{sec:singlepair}

We begin by analyzing the observables in a single-ancilla setup of Fig.~\ref{fig:diag}. Starting from the domain-wall state, we observe a nonmonotonic dependence on the coupling constant $M$ for both the domain-wall melting rate and the rate at which the entanglement grows. This makes evident a competition between different measurement-induced effects. 

For $M=0$ the dynamics is integrable, and the domain wall displays relatively slow melting dynamics, as shown for reference in Fig.~\ref{fig:M0} in Appendix \ref{sec:appendix_unc}. This behavior is similar to that studied in Refs.~\cite{ljubotina17,Misguich2017a,ljubotina19,Wei2022a} and thought to belong to the Kardar-Parisi-Zhang class (see Refs.~\cite{ljubotina19,Wei2022a} and references therein). 

\subsubsection{Effect of measurements on averaged properties}

As $M$ is increased, the melting dynamics first accelerates as a result of integrability breaking induced by measurements, which is exemplified by the case of $M=1$ in Fig.~\ref{fig:ms8L16}(a). Indeed, we observe that the number of particles transported through the center of the system has reached a value of almost $L/4$ for $t = 200$, indicating the almost complete melting of the domain wall. Transport is therefore \emph{faster} than in the uncoupled case $M = 0$.
Repeated measurements on the ancilla for $M=1$ also increase the entanglement growth in the system quenched from the domain-wall state, compared to the case \mbox{$M = 0$} 
[see Fig.~\ref{fig:ms8L16}(c)]. 
Recently, a similar measurement-induced enhancement of the entanglement growth in a Hamiltonian system was also found in the quantum Ising chain \cite{Biella2021a}. Note that the acceleration of the entanglement growth by switching on coupling of the main chain to ancilla at $M=1$ is due to the effect of repeated measurements, but not the effect of coupling as such, see Appendix \ref{aC}.

Generally, in the limit of large $M$, strong coupling between the main chain and the ancilla overcomes the effect of integrability breaking and, as already discussed in Sec.~\ref{sec:rabi}, brings dynamics to a stop. Domain-wall melting is arrested in this limit, and the ancilla boson is prevented from tunneling to the upper ancilla site by forming a repulsively bound pair with the boson on the measured site in the main chain. Projective measurements of the ancilla site only very rarely destroy this correlated state (see discussion below). A trend towards the freezing of the system dynamics is apparent by comparing the cases of $M=1$ and $M=5$ in Fig.~\ref{fig:ms8L16}. Dynamics slows down for $M=5$ compared to $M=1$, as is attested by the results in Figs.~\ref{fig:ms8L16}(a), \ref{fig:ms8L16}(b), and \ref{fig:ms8L16}(c) for the particle density in the main chain, the particle density in the ancilla, and the entanglement entropy [see the thick black lines in Fig.~\ref{fig:ms8L16}(c) that depict the average over 40 quantum trajectories] respectively. When $M$ is increased to $M = 50$, we find no appreciable melting of the domain wall or growth of the entropy (see Appendix \ref{sec:appendix_dwall}).

Aside from the general trend of, first, an increase in the melting rate as $M$ is increased to $M \approx 1$, and, second, a slowing down as $M$ is further increased to $M \gg 1$, the dependence on $M$ is more intricate. Specifically, dynamics of the system shows oscillatory behavior as a function of $M$ between $M\sim 1$ and $M\to\infty$. This is because of a possible commensurability of the Rabi frequency in the ancilla and the measurement rate $2\pi/\Delta T$. Although the density oscillations in the ancilla for a given $M$ are not exactly periodic, their spectral weight is peaked around a certain frequency, the existence of which derives from the static mean-field approximation discussed in Sec.~\ref{s2}. 
One of the consequences of this effect is an enhanced likelihood of obtaining the same result for two consecutive projections if $\Delta T/2\pi$ is close to an integer of the inverse peak frequency. For instance, under this condition, if a projection in the ancilla yields 1 for the occupation of the lower ancilla site, then the Born rule for the next projection most likely also yields 1 (quantum Zeno effect). Thus, when the resonant condition is closely satisfied, the effective measurement strength is enhanced, suppressing dynamics. This is the case for $M=3$, as seen from Fig.~\ref{fig:ms8L16}, where dynamics for all three observables ($n$, $n_a$, and $S$) is distinctly slower for $M=3$ than for either $M=1$ or $M=5$.

\subsubsection{Fluctuations: trajectory branching and the quantum Zeno valve effect}

The mechanism by which the dynamics of the entanglement entropy $S$, averaged over realizations, is suppressed with increasing $M$ reveals itself when we study the evolution of the entropy for particular quantum trajectories. These results are shown in Fig.~\ref{fig:ms8L16}(c). For $M=1$, the picture of $S$ resolved with respect to realizations is relatively featureless. 
For individual quantum trajectories, the entropy grows gradually, without sharp acceleration at any point in time, and closely approaches the thermal value (the Page entropy) already on a timescale on the order of $10^2$. The distribution of $S$ over different quantum trajectories for $M=1$ is  rather sharply peaked at any given $t$, with fluctuations becoming smaller and smaller as $S$ approaches the Page value. This forms a well-defined reference curve for the \mbox{entangling} (``thermalized'') behavior of the entropy.

Fluctuations between different realizations grow as a function of $M$. Of particular interest is the manner in which the fluctuations become more prominent for $M=3$, when the system, as discussed above, exhibits a resonance between the Rabi frequency in the ancilla and the measurement rate. As a striking feature of the realization-resolved entropy for $M=3$, a given quantum trajectory exhibits a slow growth of $S$, identical to many other trajectories, until the trajectory suddenly ``branches off''---this process is visualized as kinks on the trajectories in Fig.~\ref{fig:ms8L16}(c)---and starts quickly trending towards a highly entangled state that evolves similar to a typical state for $M = 1$. The delay time $t'$ at which the branching-off from the least entangled state occurs for $M=3$ is seen to be widely spread in Fig.~\ref{fig:ms8L16}(c) within the interval of observation.

The effect of quantum-trajectory branching can be understood by considering the probability of measuring a hole on the ancilla site. Since the system for $M=3$ is close to the resonance (see Sec.~\ref{sec:rabi} above) and the ancilla is prepared at $t=0$ in a state with $n_a=1$, ancilla projections under these conditions measure a particle during a long sequence of measurements. This explains the relatively slow (compared to the case $M=1$) growth of $S$ for time scales much larger than $\Delta T$. 
Once a hole is measured, which is a rare event near the resonance, the domain wall starts to melt much more rapidly, which triggers the rapid growth of the entanglement entropy. There hence emerges a valve effect (``quantum-Zeno-valve effect"), in which the measurement of a hole moves the system away from the resonance and the system never evolves back towards it. The quantum trajectories eventually all end up in a thermal state once they have moved away from the resonance.

The sharp branching-off of quantum trajectories, associated with the measurement of a hole in the ancilla, is in fact a more robust feature of the entanglement spreading in the setup with a domain wall, not necessarily related to the commensurability of the Rabi frequency and the measurement rate. Namely, it is a feature of the entanglement dynamics also for arbitrary $M\gg 1$. In this limit, however, the mechanism of branching is different compared to the resonant case. The amplitude of Rabi oscillations given by Eq.~\eqref{eq:rabiamp} is for $M\gg 1$ strongly suppressed, so that the likelihood of measuring a hole is small irrespective of the resonant condition. This is illustrated in Fig.~\ref{fig:ms8L16}(c) by the behavior of individual quantum trajectories for $M=5$, which is qualitatively similar to that for $M=3$. Note, however, that for $M = 5$ a larger fraction of trajectories have moved towards a thermal-type state, compared to the resonant case $M = 3$.

The evolution of the behavior of $S$ for individual trajectories as a function of $M$ involves an interplay of the above two mechanisms of branching, which leads to a nonmonotonic dependence of a typical delay time $t'$ on $M$. Namely, for $M=5$, it is much larger than for $M=1$, but substantially smaller than for $M=3$. More specifically, the distribution of $t'$ over trajectories for given $M$ is parameterized in Fig.~\ref{fig:ms8L16}(c) by plotting the evolution of $S(t)$ for a series of representative trajectories, sorted by the entropy integrated over time within the interval of observation,
\begin{equation}
\Xi = \int_0^{t_\mathrm{f}} \mathrm{d}t \, S(t), \label{eq:entang_sorting}
\end{equation}
for the final time $t_\mathrm{f} = 200$. For each $M$, we have shown the curves having the highest $\Xi$ (purple line), the lowest (blue), and the median (green). We have also plotted the curves that correspond to $\Xi$ higher than 25 (orange) and 75 (red) percent of curves respectively.

\begin{figure*}[!htb]
    \centering
    \includegraphics[width=\textwidth]{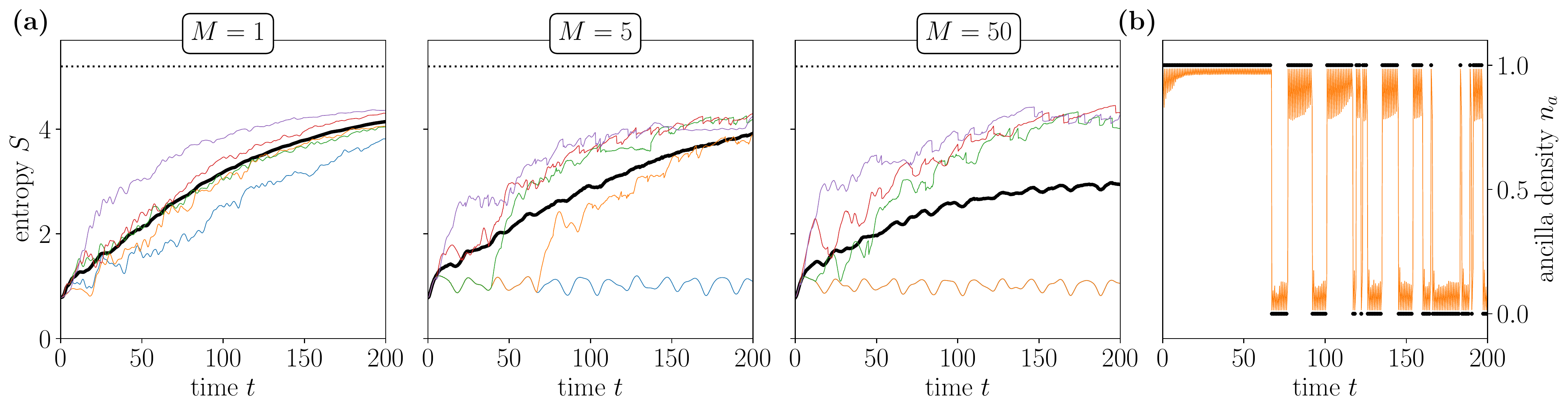}
    \caption{\textbf{(a)}: Von Neumann entropy of entanglement $S$ for the case of a single ancilla coupled to the chain (at the site $i = L/2$) initialized in the ground state, as a function of time for various $M$ and $R = 40$ realizations. The parameters are $L=16$, $U=1/4$, $\Delta T=1$. The meaning of the different curves is as in Fig.~\ref{fig:ms8L16}c. Note that in the case $M=50$ (right panel) the lowest-entropy curves (orange and blue) overlap. \textbf{(b)}: Dynamics of the ancilla density $n_a$ corresponding to the trajectory shown by the orange curve in the panel $M = 5$. The black dots represent measurement outcomes $\{0, 1 \}$.}
    \label{fig:dmrg_gs_singleanc}
\end{figure*}

For $M = 1$, the characteristic $t'$ is of the order of unity, but the delay is seen to dramatically increase by two orders of magnitude to $t'\sim 10^2$ for $M = 3$, with the median trajectory (green curve) branching off at $t'\approx 175$. No branching whatsoever occurs for $M=3$ within the interval of observation for the least entangled trajectories in the bottom 25th percentile. The fact that the typical delay time is so large for $M=3$ attests to the resonance as the prime reason for the delayed branching. For $M = 5$, the delay time for the median trajectory decreases, compared to $M=3$, to $t'\approx 60$, pointing to the suppression of Rabi oscillations as a mechanism of the delay. For very strong coupling, $M = 50$, no branching at all is observed within the time $t_\text{f}=200$, see Fig.~\ref{fig:domainwall_M50} in Appendix \ref{sec:appendix_dwall}.

\subsection{Ground state as the initial state} 
\label{sec:dmrg_gs}

In Sec.~\ref{sec:singlepair}, we have considered the case of a single ancilla coupled to the main chain, initialized in a domain-wall state.
Let us now turn to the case where the initial state is the ground state of the uncoupled system ($M=0$) with $U = 1/4$, as schematically depicted in Fig.~\ref{fig:diag_dmrg} for two ancillas. This initial state is homogeneous (aside from boundary effects), features power-law correlations of the Luttinger-liquid type, and a finite entanglement entropy that depends logarithmically on $L$ \cite{Laflorencie2016a}. A useful feature of this choice is that measurement-induced disentangling behavior of the system can be probed more easily, as the entanglement entropy then decreases with respect to that in the initial state \cite{Doggen2022a}. Conversely, we shall find that the quantum-Zeno-valve effect is less pronounced in this case, since domain wall melting is initially constrained by dynamics at the domain wall only, while particles can more easily rearrange themselves starting from the half-filling delocalized ground state. This means that the dynamics also more easily moves away from the resonant condition as discussed in Sec.~\ref{sec:rabi}.

\subsubsection{Single ancilla}

\begin{figure*}[ht]
\centering
\includegraphics[width=.75\textwidth]{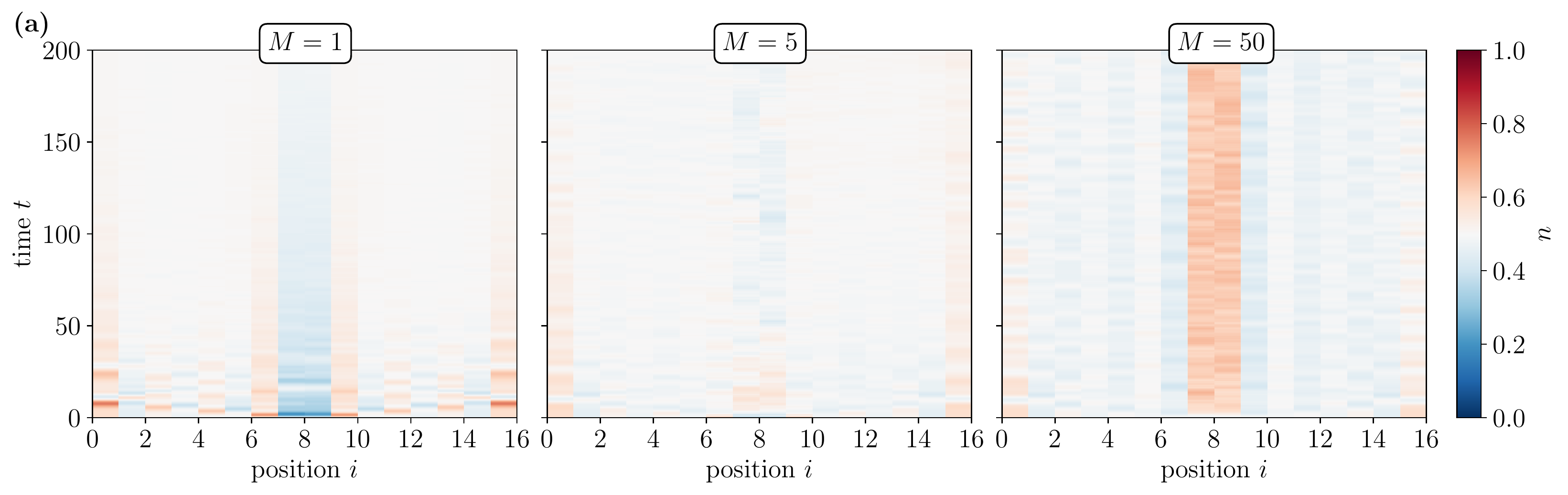}
\includegraphics[width=.24\textwidth]{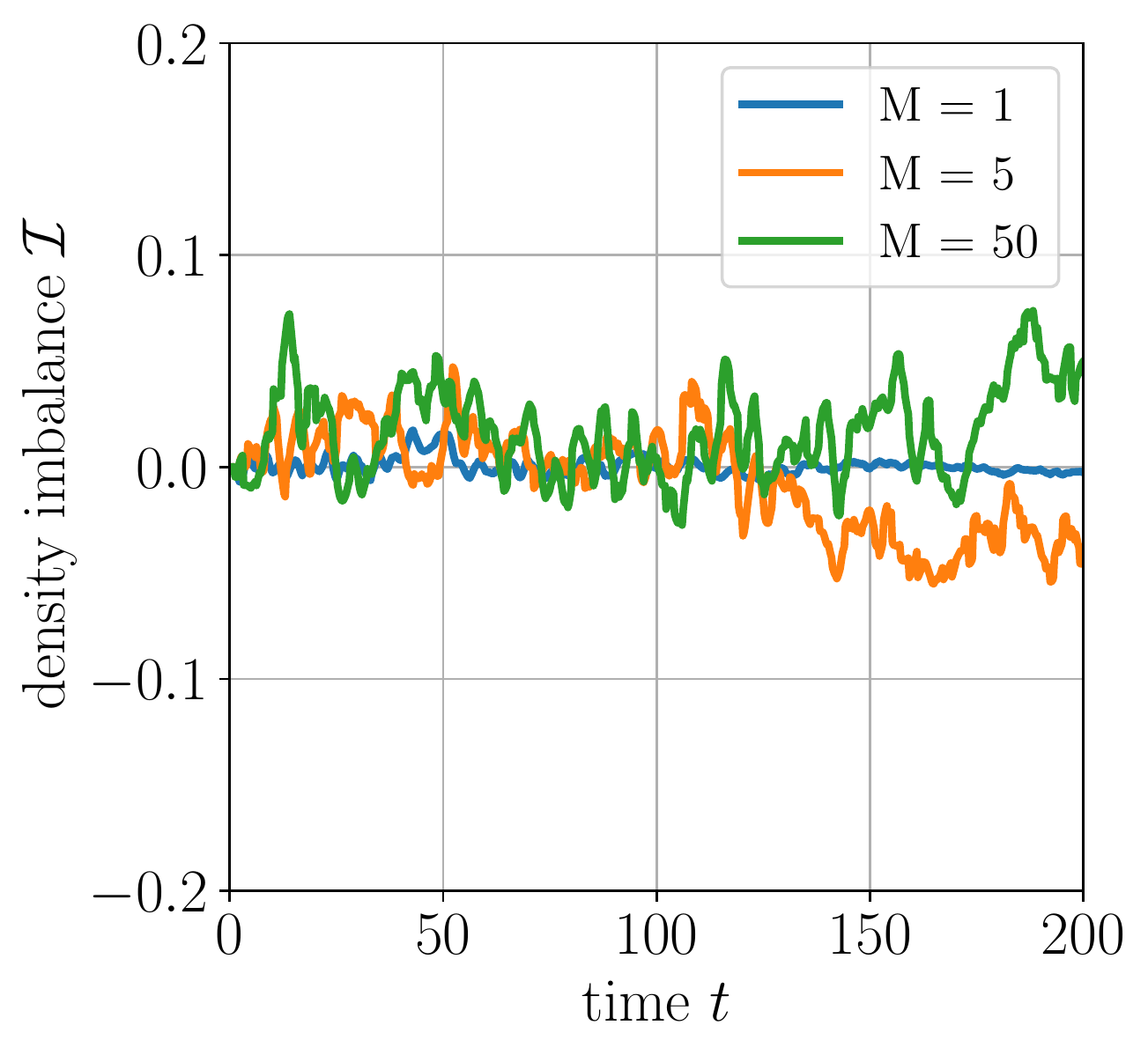}
\includegraphics[width=.275\textwidth]{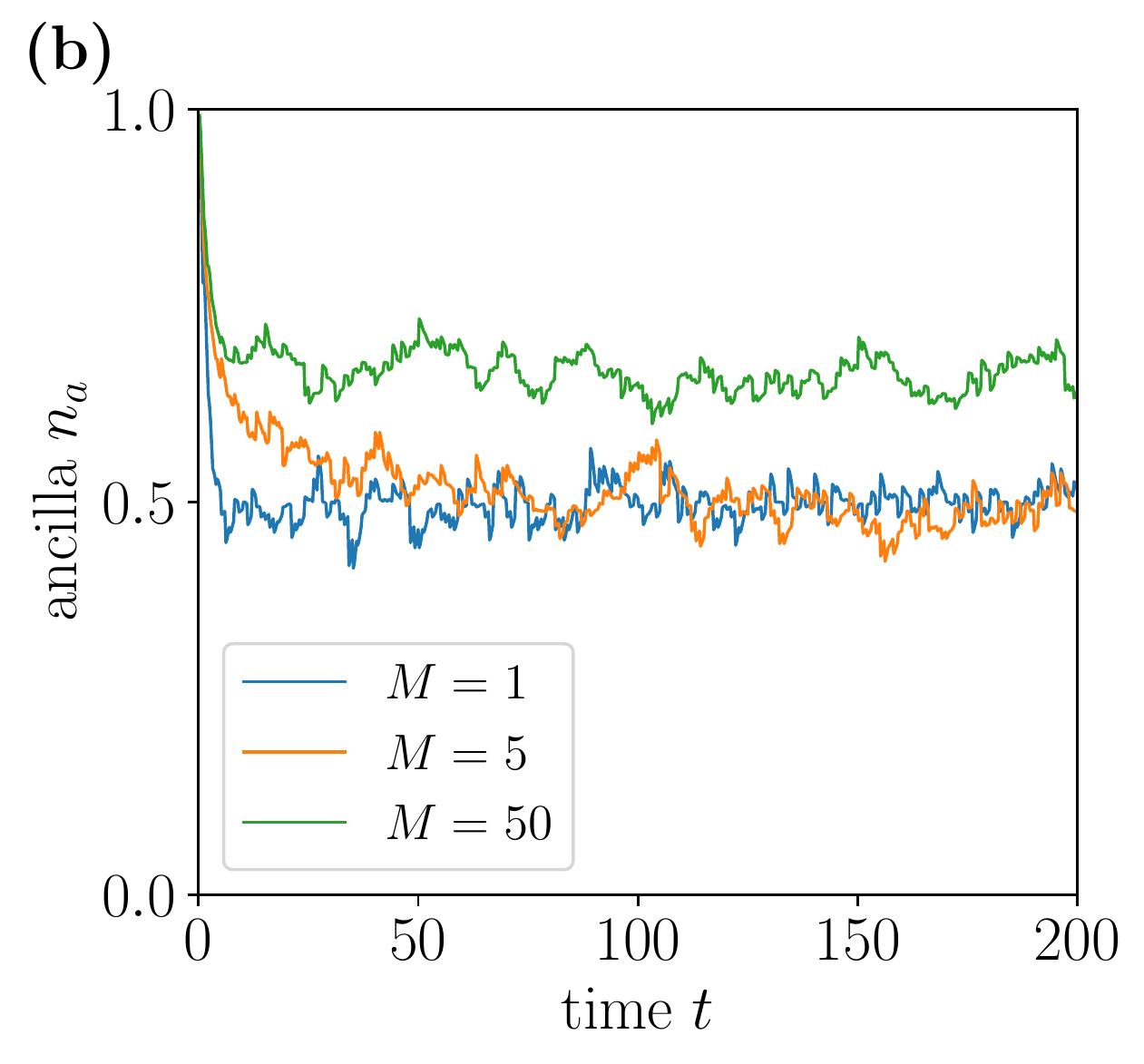}
\includegraphics[width=.715\textwidth]{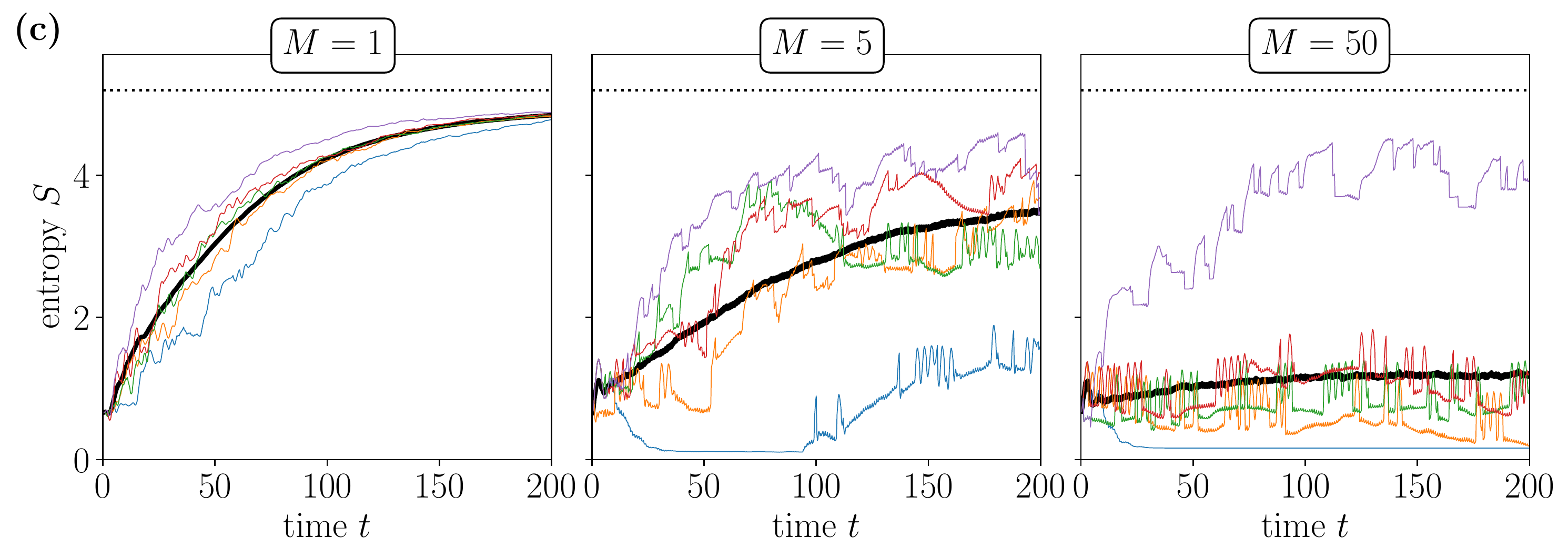}
\caption{Dynamics of the system with two ancillas and the chain initially in the ground state, Fig.~\ref{fig:diag_dmrg}, for $L = 16$, $U=1/4$, and $M=1,5,50$. The measurement period $\Delta T = 1$. \textbf{(a)}: Particle density $n_i(t)$ in the main chain as a function of time and the associated bipartite particle fluctuations indicated by the density imbalance $\mathcal{I}$. \textbf{(b)}: Particle density $n_a$ of the left ancilla as a function of time, averaged over about 300 realizations. \textbf{(c)}: von Neumann entropy of entanglement $S$ as a function of time. All curves in (c) have the same meaning as in Figs.~\ref{fig:ms8L16}(c) and \ref{fig:dmrg_gs_singleanc}.}
\label{fig:ms4L16dmrg}
\end{figure*}

In Fig.~\ref{fig:dmrg_gs_singleanc}a, we show results for the entropy dynamics for a single ancilla, when the main chain is initialized in the ground state. 
In general, the qualitative behavior of the entropy is analogous to the case of the domain-wall initial condition (Sec.~\ref{sec:singlepair}), see Fig.~\ref{fig:ms8L16}(c).
For $M=1$, the entropy quickly grows with relatively small  fluctuations that decrease with time. For intermediate coupling, $M=5$, the quantum-Zeno-valve effect is apparent in the form of a sharp branching-off from the low-entanglement curve with a broad distribution of time $t'$. 

It is instructive to also consider the ancilla density $n_a$ for the case when the the quantum-Zeno-valve effect is clearly visible. In Fig.~\ref{fig:dmrg_gs_singleanc}b we show $n_a$ as a function of time for a trajectory indicated with the orange curve in panel \textbf{(a)} for $M = 5$. We see that the ancilla density remains close to unity, with measurement outcomes unity, up to $t' \approx 70$, a manifestation of the quantum-Zeno-valve effect. During this time, the entropy also remains close to its initial value. After the rare event of measuring a hole in this nearly frozen state (which happens at $t' \approx 70$ for this trajectory), the measurement results start to jump between zero and unity since the state is no longer stabilized by a bound pair. This leads to branching in the entanglement entropy curve and to thermalization.

For very strong coupling, $M=50$, we observe a clear bimodal distribution of the entropy. Specifically, for about half of the quantum trajectories, the system quickly ``thermalizes,'' whereas the other half of trajectories stay at low entanglement close to that of the initial state. 
This behavior can be understood as follows. 
The initial quench---coupling the ancilla to the chain---introduces high energy $M$ in the components of the wave function with $n_{L/2}=1$. After projecting the ancilla, this component survives in about half of trajectories. After this, a blockade occurs for these trajectories, in a full analogy with the case of the domain wall (where $n_{L/2}=1$ initially for all the trajectories), see Sec.~\ref{sec:rabi} and Fig.~\ref{fig:domainwall_M50}. The stability of such a ``repulsive bound state'' formed on the ancilla site and the chain site with $i=L/2$ is explained by the energy conservation constraint. Indeed, this highly excited state cannot easily decay, since $M$ is much larger than the bandwidth.  

We conclude that for a single ancilla with sufficiently large $M$, independently of the initial state, thermalization is accompanied by a formation of a bimodal distribution of quantum trajectories. The trajectories show sharp branching-off from the low-entropy to the high-entropy class of states, with a broad distribution of branching times $t'$.

\subsubsection{Two ancillas}

With a view to ultimately connecting to the physics behind measurement-induced transitions \cite{Skinner2019a,Li2018a} and as a step towards the finite density of measured sites (Sec.~\ref{sec:5D}), we enhance the effect of measurements by considering now two ancilla pairs, one connected to site $L/2$ and the other to site $L/2 + 1$ in the middle of the chain (see Fig.~\ref{fig:diag_dmrg}). This setting also allows us to explore possible correlations, mediated by the dynamics of the main chain, between multiple ancillas. 
The measurement part of the Hamiltonian is a straightforward generalization of Eqs.~(\ref{eq:meas_ham}) and (\ref{eq:meas_ham_sa}) with identical interaction ($M$) and hopping ($J$) constants for both ancilla pairs.

In the measurement protocol, a projective measurement is performed on the left ancilla pair ($i = L/2$) first, followed after an infinitesimal time interval by a projective measurement on the right one ($i = L/2 + 1$). In a sequence of repeated measurements, a ``double salvo" of such two measurements is separated from the next one by the time interval $\Delta T$.
As in the domain-wall setup, the bipartite entanglement entropy $S$ is computed with respect to partition in the middle of the main chain. Considering the quench from the ground state, we analyze statistics over a larger number of individual trajectories, namely about 300, compared to 40 in Sec.~\ref{sec:singlepair}.

The numerical results for $M = 1, 5, 50$ in the double-ancilla setup of Fig.~\ref{fig:diag_dmrg} are shown in Fig.~\ref{fig:ms4L16dmrg}. For $M=1$, the entanglement dynamics for the quench from the ground state with two ancillas is very much similar to that for a single ancilla. Specifically, the entanglement entropy grows with time in a smooth manner, gradually evolving to the thermal value, with a rather narrow distribution of the entropy over individual quantum trajectories. The shape of the typical entropy curves here is similar to the curve in Figs.~\ref{fig:ms8L16}(c) and \ref{fig:dmrg_gs_singleanc} for $M=1$. 
The trajectory-averaged ancilla density $n_a$  [Fig.~\ref{fig:ms4L16dmrg}(b)] quickly loses memory of the initial condition for $M=1$, fluctuating around the average $n_a = 1/2$, also similar to the single-ancilla case [Fig.~\ref{fig:ms8L16}(b)].

When coupling $M$ becomes stronger, $M=5$, the distribution of entropy broadens in analogy with the case of a single ancilla. 
At the same time, an essential difference is observed. Specifically, the entropy growth for individual quantum trajectories does not show sharp branching between the two well-defined types of behavior described by the low- and high-entanglement curves (compare the panel for $M=5$ in 
Fig.~\ref{fig:ms4L16dmrg} with those in
Fig.~\ref{fig:ms8L16} and Fig.~\ref{fig:dmrg_gs_singleanc}). Instead, the growth follows individual curves that are characterized by a rather broad distribution without apparent bimodal features. This can be attributed to the mutual ``interference'' effect of the two ancillas, tending to blur the quantum-Zeno-valve phenomenon.

\begin{figure}
\centering
\includegraphics[width=\columnwidth]{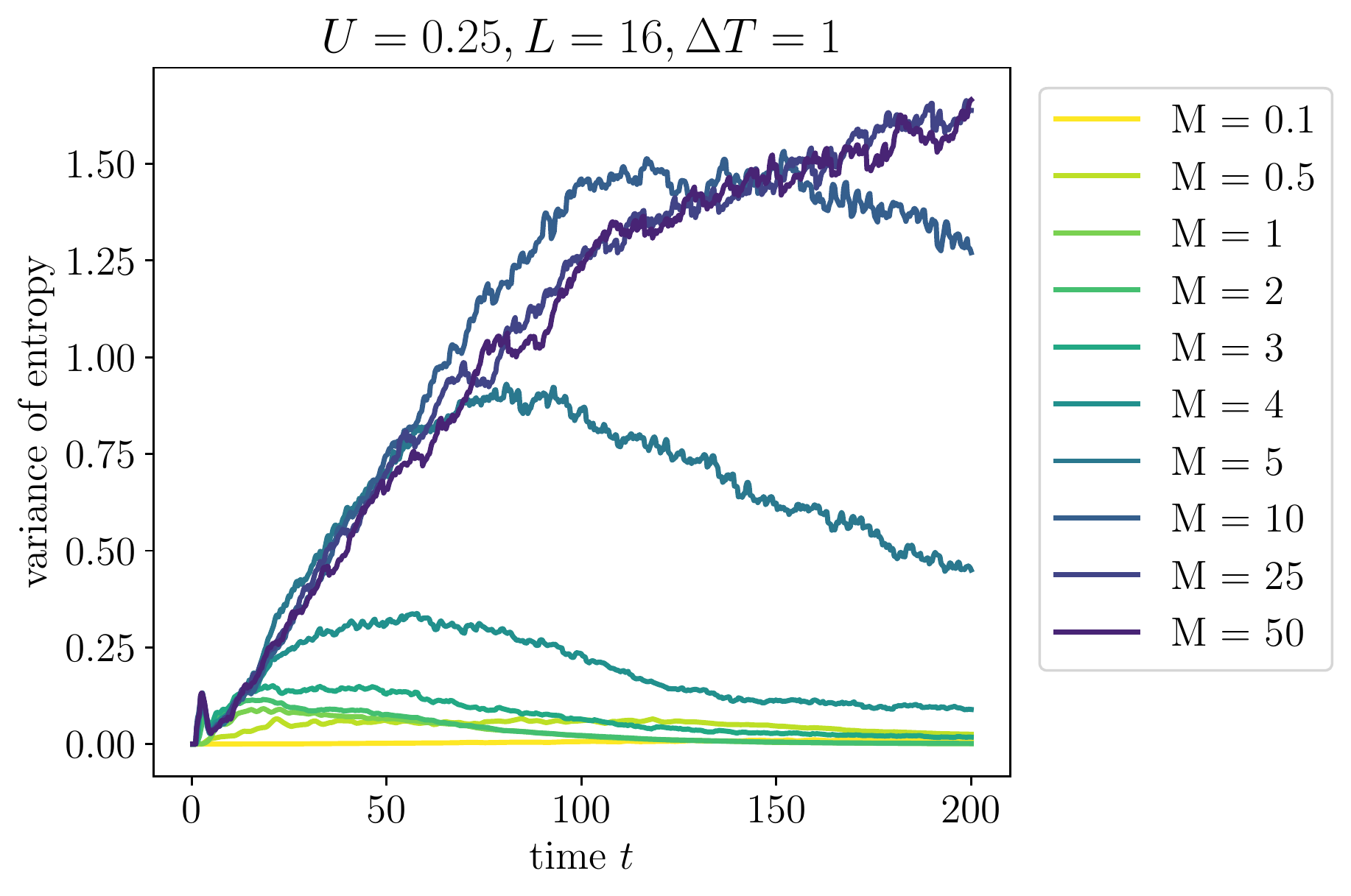}
\caption{Variance of the entanglement entropy across $R \approx 300$ different realizations in the setup with two ancillas (Fig.~\ref{fig:diag_dmrg}) as a function of time, for various values of $M$ and the same $L$, $U$, and $\Delta T$ as in Fig.~\ref{fig:ms4L16dmrg}.}
\label{fig:variance}
\end{figure}

\begin{figure*}
\centering
\includegraphics[width=.85\textwidth]{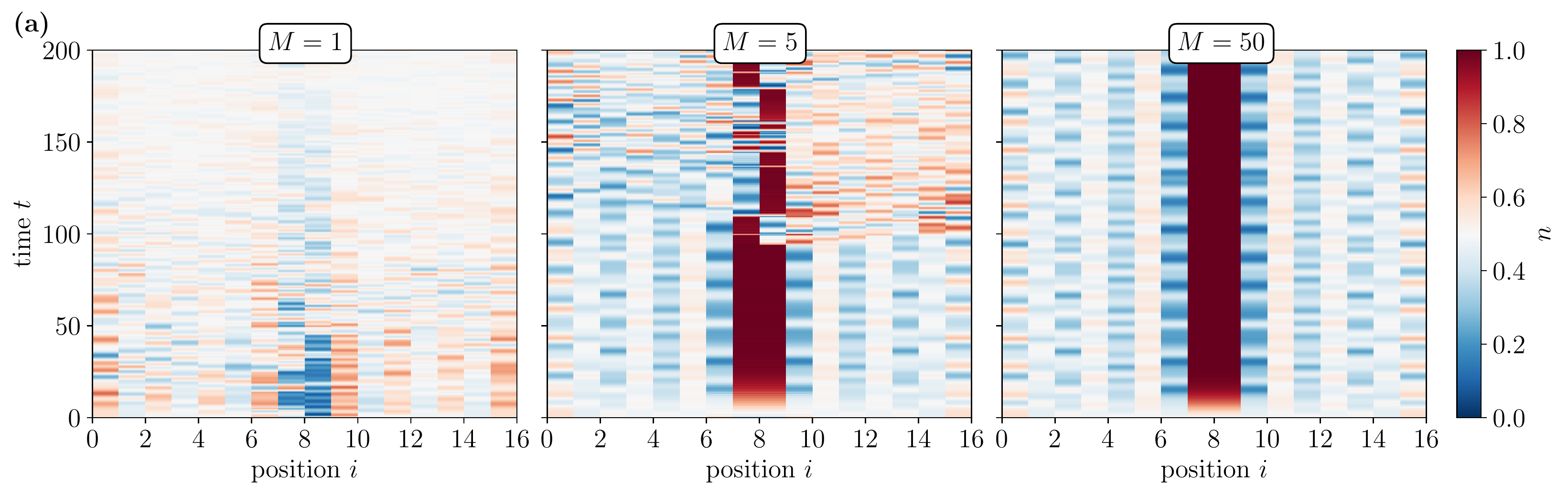}
\includegraphics[width=.85\textwidth]{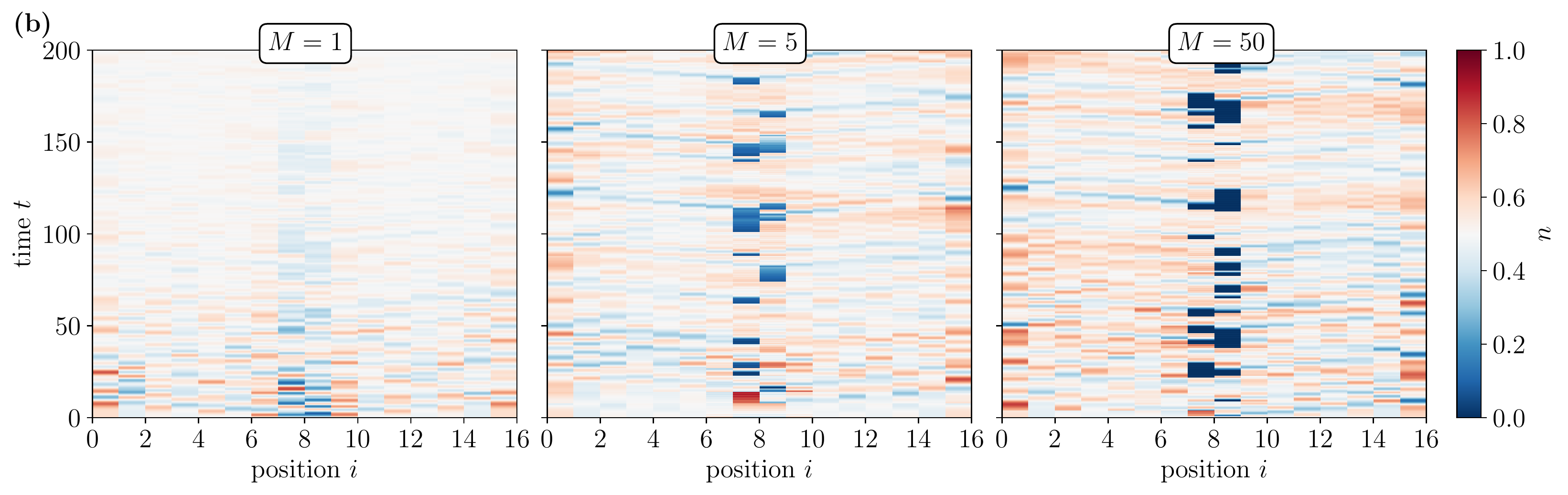}
\caption{Density in the main chain for quantum trajectories corresponding to the \textbf{(a)} smallest and \textbf{(b)} largest integrated entanglement entropy $\Xi$ [Eq.~(\ref{eq:entang_sorting})] for various $M$ in the double-ancilla system described in Fig.~\ref{fig:ms4L16dmrg}.}
\label{fig:ms4L16indiv}
\end{figure*}

Another striking difference is the emergence of up-and-down jumps, about 0.5--1 in magnitude, visible in the quantum trajectories of the entropy for $M=5$ and $M=50$. They are the signature of measurement-induced fluctuations of the particle number in halves of the chain, as is evidenced by a comparison between the timing of the jumps in the entropy and the timing and position of the density fluctuations. Note that the size of the jumps is close to the contribution to the entanglement entropy associated with a single Bell pair: $\ln 2$.

For $M=50$, the entropy behaves similarly to the single-ancilla case in that there are two classes of trajectories (with low and high entanglement). However, there are also clear differences. 
First, a smaller fraction of trajectories features high entanglement. This is because the blocking by two ancillas is more efficient, since it is operative when at least one particle occupies sites $L/2$ and $L/2+1$. 
Thus, the averaged entropy is strongly suppressed for the case of two ancillas. This provides indication for a formation of a disentangling phase for many ancillas, as discuss below in Sec.~\ref{sec:5D}.

To complete the statistical description of quantum trajectories, we consider the distribution of the entanglement entropy as a function of time. We characterize the distribution by the variance:
\begin{equation}
\mathrm{var}_S(t) = \frac{1}{R} \sum_{j=1}^R \left[S_j(t) - S_\mathrm{avg}(t)\right]^2,
\end{equation}
where $R$ is the number of realizations and $S_\mathrm{avg}$ is the mean value of the entropy. The result is shown in Fig.~\ref{fig:variance}, where a strong dependence on the coupling strength $M$ is evident, with a crossover around $M=4$ from a narrow distribution to a wide one. The fluctuations of $S$ are nonmonotonic in time, with an initial increase of the distribution width followed by a decrease, except for the largest values of $M$. The narrowing of the distribution can be attributed to more trajectories converging to the thermal state for large $t$. For $M=25$ and 50, this narrowing is not seen because only a few rare trajectories are strongly entangled.

\subsubsection{Density distribution in the main chain for two ancillas}

In Fig.~\ref{fig:ms4L16dmrg}(a), we show the averaged density in the chain as a function of time. For $M=1$, we observe an initial suppression of the density at the sites coupled to the ancillas. This inhomogeneity is faded away with time. For strong coupling, $M=50$, a very different behavior is observed: the density is enhanced and this enhancement stays approximately constant within the observation time. This is a manifestation of a formation of a repulsive bound state, discussed above (for a single ancilla).

To further illustrate this, we present the density dynamics on the least and the most entangled trajectories for various $M$ in Fig.~\ref{fig:ms4L16indiv}. 
The panels for the least entangled states at $M=5$ and $M=50$ in Fig.~\ref{fig:ms4L16indiv}(a) clearly demonstrate the emergence, as $M$ is increased, of a stable blocking region, with at least one of the two sites connected to the ancilla staying occupied. By contrast, the most entangled trajectories feature chaotic dynamics with frequent occurrence of holes on these sites. Recall that holes do not interact with the ancilla in Eq.~\eqref{eq:meas_ham_sa}, whereas a particle on the site coupled to the ancilla is pinned by strong interaction with the ancilla particle, leading to a repulsively bound pair that is only broken once a hole is measured in the ancilla. This is much the same physics as pertinent to freezing-out of the system with the domain-wall initial state in the limit of large $M$. 

We also observe that the dynamics of the density is correlated with the entanglement dynamics. As an example, 
the one-to-one correspondence between the entropy jumps and the fluctuations of the particle number imbalance can be clearly seen by comparing the evolution of $S$ on the least entangled trajectory for $M=5$ in Fig.~\ref{fig:ms4L16dmrg}(c) and the density fluctuation for $M=5$ on this trajectory in Fig.~\ref{fig:ms4L16indiv}(a). The presence of spikes in the entropy curves for the double-ancilla setup, as opposed to the case of a single ancilla, again indicates the importance of correlation in the dynamics of multiple ancillas caused by their interaction with the same main chain in the course of joint unitary evolution.

\section{Measurements on every site}
\label{sec:5D}

With the above detailed analysis of the previous cases with one and two ancillas, we are now in a position to proceed with the setup where each site of the main chain is coupled to an ancilla pair.

\begin{figure*}[!htb]
    \includegraphics[width=\textwidth]{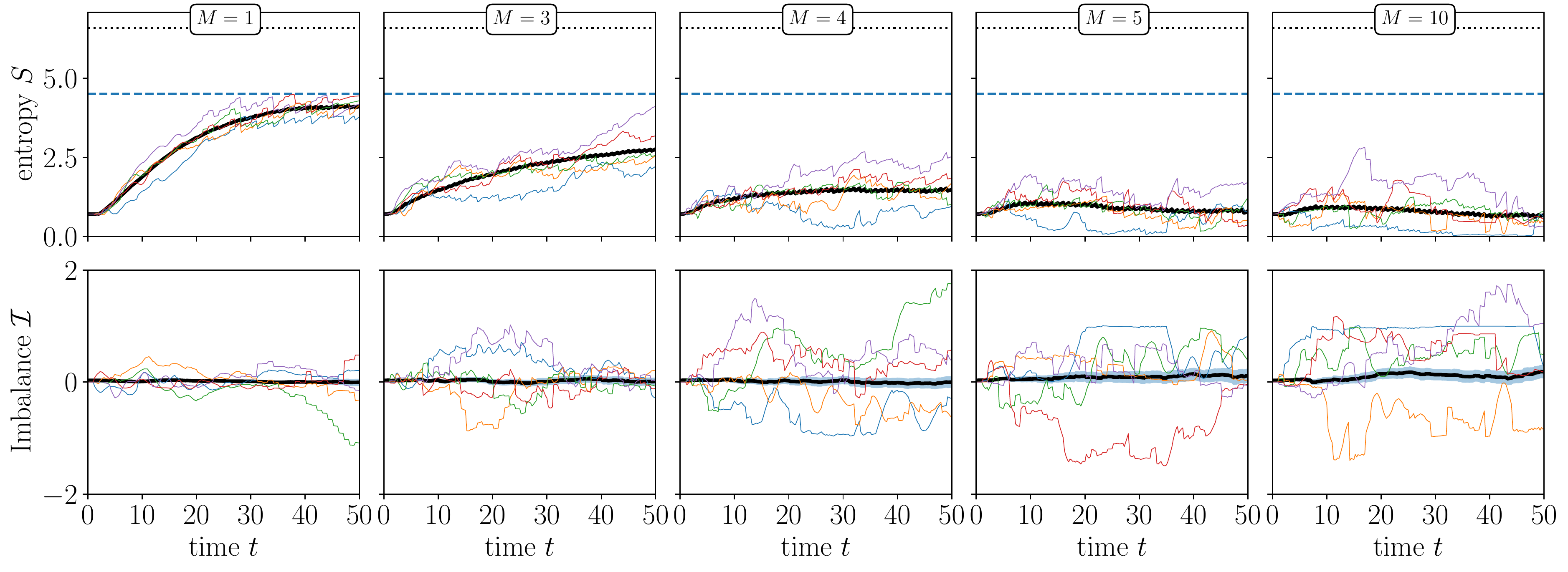}
    \caption{Dynamics for the many-ancilla case schematically depicted in Fig.~\ref{fig:projdiag_multi}, with $U = 1/4$, $L = 20$, and $\Delta T = 1$. Top panels: average entropy (thick black line) over $R = 160$ trajectories for various values of $M = 1, 3, 4, 5, 10$. Colored lines indicate trajectories corresponding to various values of the integrated entropy $\Xi$ [Eq.~(\ref{eq:entang_sorting})], as defined in Fig.~\ref{fig:ms8L16}(c). The dotted black line indicates the Page entropy, while the dashed blue line indicates the maximum entropy that can be obtained with the chosen bond dimension $\chi = 128$. Bottom panels: particle imbalance \eqref{eq:imba} as a function of time. The color-coded lines correspond to the same trajectories as in the top panels. The shaded area around the average denotes a $2\sigma$ confidence interval.}
    \label{fig:manyanc_diffM}
\end{figure*}


\subsection{Measurement protocol}

The consideration of a finite density of ancillas allows us to study the ancilla-measurement approach in the context of entanglement transitions. As in Sec.~\ref{sec:dmrg_gs}, we initialize the main chain in the ground state with $U = 1/4$. We address system sizes up to $L = 24$, so that the total number of sites in the system is $L_\mathrm{tot} = 24 \times 3 = 72$. This means that we can no longer use an unrestricted bond dimension $\chi$ (i.e., exact numerical simulations) as before. The bond dimension controls the size of the variational subspace considered during time evolution \cite{Haegeman2016a, Schollwock2011a}. We furthermore restrict our time window to $t \in [0, 50]$ as the simulations become more computationally demanding.

The measurement protocol is now as follows. At each step $\Delta T$, we perform projective measurements on all ancilla sites connected to the main chain, using the Born rule as before. These measurements are performed consecutively from left to right, so that each ancilla is projected at every measurement cycle, with an infinitesimal delay between the measurements on
neighboring sites.

\subsection{Overview of entanglement dynamics}

In the double-ancilla case, we observed that having two ancilla pairs in the system is not sufficient to reach a disentangled state for every trajectory, even for very large measurement strength $M = 50$. It is therefore of interest to see whether a disentangling phase forms in the limit of a large number of ancilla pairs. Furthermore, it is even more intriguing questions whether our setup yields an entanglement transition and what are then the properties of the entangling phase. In particular, while we observed a thermalizing (reaching nearly maximum entanglement) behavior for one and two ancillas with $M=1$, it is not obvious whether this behavior survives in the case of all sites coupled to ancillas.

In Fig.~\ref{fig:manyanc_diffM}, we show the resulting dynamics of the entanglement entropy 
for the chain length $L=20$ and various choices of $M = 1, 3, 4, 5, 10$.
For all values of $M$ we see an initial growth of $S$ at $t\lesssim 10$, induced by the quench. For longer times the dynamics for different couplings is very different. 
For large measurement strength, $M = 10$, we observe that entanglement is suppressed on average, as shown by the thick black line, compared to entanglement in the initial state. The most entangled trajectory as quantified by $\Xi$ (thin purple line), unlike the double-ancilla case, also exhibits suppressed entanglement. We have verified (result not shown) by comparing the cases of bond dimensions $\chi=128$ and $\chi = 256$ that the value of the average entropy in the large-$M$ limit ($M \geq 5$) is not dependent on the cutoff determined by the bond dimension. 

For sufficiently weak measurements, $M = 1$, we observe a fast growth of entanglement with a quick saturation at the cutoff value. The entropy follows the curve that is almost identical in shape to those for the single- and double-ancilla setups at $M=1$, which approached the Page value proportional to the system size (see Sec.~\ref{sec:repmeas}). Therefore, we expect that in this case the exact dynamics (which is numerically inaccessible for long times $t\gtrsim 15$ due to strong entanglement growth) should closely approach the Page value indicated by the dotted black line. The clear difference in the entanglement dynamics for $M=1$ and $M=10$ strongly suggests an existence of a measurement-induced entanglement transition in between. 

\begin{figure*}[!htb]
    \centering
    \includegraphics[width=\textwidth]{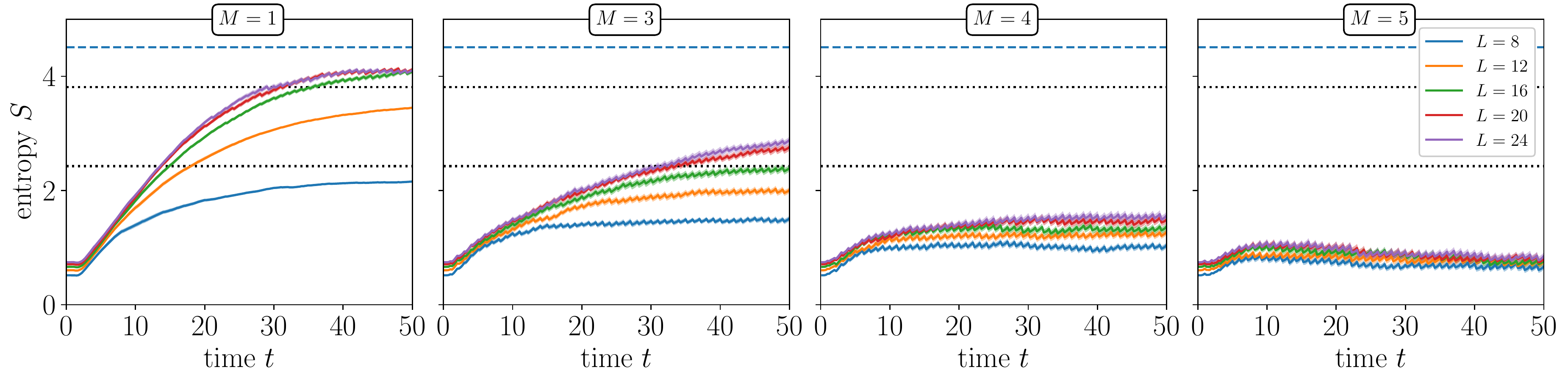}
    \caption{Trajectory-averaged dynamics of the entropy for various choices of $M = 1, 3, 4, 5$ and system sizes $L = 8, 12, 16, 20, 24$. The number of trajectories $R = 160$. The shaded area around the curves indicates a $2\sigma$ confidence interval. Horizontal black dotted lines indicate the Page entropy for $L = 8$ and $L = 12$ respectively. The dashed line is the entropy cutoff due to the finite bond dimension $\chi = 128$.}
    \label{fig:manyanc_diffL}
\end{figure*}

Let us now inspect the entanglement dynamics for the intermediate values of $M$. The dynamics for $M = 5$ is very similar to the case $M = 10$, indicating that $M=5$ also belongs to the disentangling phase. For $M=4$, we observe a very weak increase of the entropy, which is followed by a saturation at a value that is only slightly larger than the initial one.  
For $M = 3$, there is a clear trend of increasing entropy with time, although the increase is much slower than for $M=1$.
At this value of $M$, we observe a dependence on the bond dimension $\chi$ at later times.
This consideration suggests that the entanglement transition takes place in the vicinity of $M=4$. To shed more light on this, we will consider the $L$-dependence of the entropy in Sec.~\ref{sec:Lentropy}. 

For all values of $M>1$ in Fig.~\ref{fig:manyanc_diffM}, we see a broad distribution of the entanglement entropy over quantum trajectories. 
At the same time, no bimodality features are observed in these distributions. We argue that the bimodality that we encountered in the single- and two-ancilla cases (Sec.~\ref{sec:repmeas}) is smeared by the interference of multiple ancillas in our implementation of measurements.

\subsection{System-size dependence of entropy and measurement-induced transition}
\label{sec:Lentropy}

In Fig.~\ref{fig:manyanc_diffL}, we investigate the disentangling-entangling transition more closely by showing the behavior of the entropy as a function of time for various system sizes $L = 8, 12, 16, 20, 24$ for $M = 1, 3, 4, 5$. 
For $M = 1$ (entangling phase) we observe a clear dependence on the system size. Furthermore, we observe a fast increase, approximately linear in time for $t \lesssim L$, of the entropy towards its maximum value. For small systems ($L=8, 12$), this maximum is determined by the Page entropy and for larger system sizes ($L \ge 16$) it is determined by the cutoff established by the bond dimension $\chi$.
This implies that in this case the exact dynamics is reproduced up to times ${t\approx 15}$ for $L \geq 16$. In view of the bond-dimension cutoff, the curves for $L =16, 20$, and 24 remain nearly identical in the whole considered time range. The saturation value increases approximately proportional to $L$ for $L= 8, 12$, and 16. This indicates a volume-law dependence of the entanglement entropy. Weak fluctuations around the average, as seen in Fig.~\ref{fig:manyanc_diffM},
like in the single- and double-ancilla setups for $M=1$ (where the entropy reached the Page value), are also suggestive of volume-law behavior. A volume law for not too large $M$ is consistent with earlier results on interacting Hamiltonian systems obtained with different implementations of measurements~\cite{Fuji2020a,Goto2020a,Doggen2022a}. 

On the other hand, for $M = 5$, the curves collapse within error bars for different $L$, a feature of area-law scaling. In the intermediate case $M = 4$, we observe a very weak growth with system size, just about distinguishable within error bars. This behavior can be potentially described as a logarithmic growth with a relatively small prefactor
(which is still larger than that in the Luttinger-liquid ground state \cite{Laflorencie2016a}). Alternatively, the weak $L$ dependence here can, in principle, be attributed to finite-size corrections convergent to an area-law curve. More work is needed to distinguish between the two possibilities.
Note that this value of $M$ corresponds to the crossover scale identified in the previous section, associated with the width of the distribution of the entropy for different trajectories in the two-ancilla case (see Fig.~\ref{fig:variance}).
For $M=3$, we observe a clear increase of the long-time value of entropy (note that the largest-$L$ curves start to be slightly affected by the bond-dimension cutoff, as mentioned above). This supports the interpretation of $M=3$ belonging to the entangling phase. 

This analysis confirms our above expectation 
that the transition is located near $M=4$. A more precise determination of the critical measurement strength would necessitate larger system sizes, as well as a longer time window for the simulations. Likewise, unambiguously distinguishing a volume-law asymptotic behavior from a possible critical phase with logarithmically growing entanglement with a large prefactor is difficult in numerical simulations in view of system-size and finite-time limitations.

\subsection{Density fluctuations}

To make a connection between the entropy and density dynamics, we also analyze the fluctuations in the particle density by considering the particle imbalance $\mathcal{I}$.
As before, this quantity measures the number of particles in the main chain that have moved from the right side of the system to the left side. We depict the dynamics in the bottom panels of Fig.~\ref{fig:manyanc_diffM}. As might be expected, these fluctuations generally increase as a function of $M$, with the average state nonetheless being close to homogeneous density. 

\begin{figure*}
    \includegraphics[width=.9\textwidth]{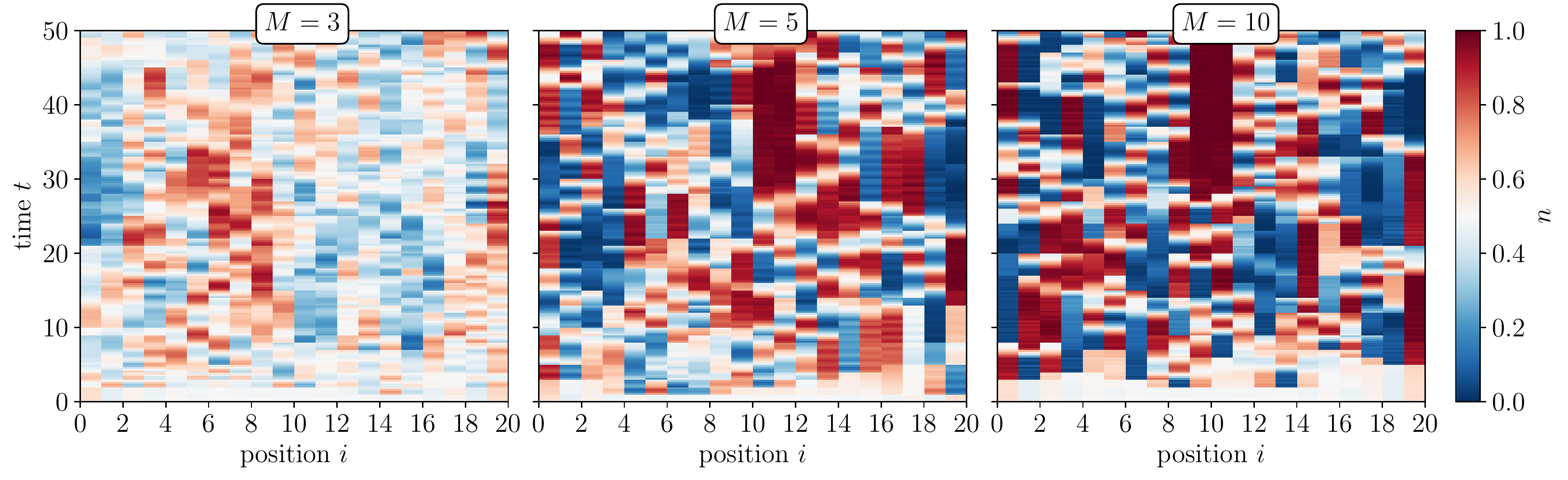}
    \caption{Density evolution along the quantum trajectories corresponding to the smallest integrated entanglement entropy $\Xi$ for various $M = 3, 5, 10$ in the many-ancilla case schematically depicted in Fig.~\ref{fig:projdiag_multi}. The system size $L = 20$ and the number of realizations $R = 160$.}
    \label{fig:manyanc_density}
\end{figure*}


For the results in the disentangling phase, with $M = 5$ and $M = 10$, we find occasional ``plateaus'' in the imbalance, associated, in particular, with weakly entangled trajectories (thin blue lines in both upper and lower panels of Fig.~\ref{fig:manyanc_diffM}). Such a plateau can be associated with weakly entangled trajectories corresponding to blocking regions in the two-ancilla case (see Fig.~\ref{fig:ms4L16indiv}a). In the many-ancilla case, such a blocking region can occur at any point in the chain. Once such a blocking region is formed, it is long-lived (for sufficiently strong coupling $M$), resembling the quantum Zeno-valve effect observed for a single ancilla. This is illustrated by considering the trajectories corresponding to the weakest bipartite entanglement, as shown in Fig.~\ref{fig:manyanc_density}.

Here we again see a clear qualitative difference between the cases $M = 3$ (entangling phase) and $M = 5$ (disentangling phase). In the former case, the density in the main chain rarely approaches the fully polarized values corresponding to a particle or hole, but shows only moderate fluctuations around the half-filling density. For $M = 5$, however, one can observe polarized regions with strongly peaked density close to 1 and 0 (dark red and dark blue, respectively). The case $M = 10$ looks qualitatively similar to $M = 5$, supporting the observation that both cases correspond to the disentangling phase. We note that in these cases, the minimum entropy trajectories feature the blocking regions at the center of the chain, where the bipartition cut is located (as in the two-ancilla setup).

The above analysis demonstrates the capability of the method to detect the measurement-induced transition in a correlated many-body system.
Our results provide further indications that measurement-induced transitions are a universal feature across a variety of measurement techniques, through employing a measurement protocol different from those conventionally used in this context. 

\section{Discussion and conclusions}
\label{s6}

In this paper, we have developed an implementation of a measurement apparatus for generalized measurements on quantum many-body systems, which is based on the coupling of the system sites to ancillary degrees of freedom that are stroboscopically projected. Our protocol can be implemented in experimental setups with, e.g., ultracold atoms, trapped ions or superconducting qubits, and can be readily adapted to generalized measurements of observables other than density by adjusting the measurement part of the Hamiltonian \eqref{eq:meas_ham_sa}.

With this tool, we have studied the effect of projective measurements of the ancillary degrees of freedom on the dynamics of hard-core bosons in the ``main'' system. Our work complements previous numerical studies in the field of measurement-induced transitions by implementing the measurements in such a way that the effect on the system of interest is indirect, through the ancilla sites that are entangled with the system. In this sense, it is a more realistic description than approaches that mimic generalized measurements or continuous monitoring through an \textit{ad hoc} stochastic formalism. We have started with the setups where the system is coupled to a single or a double ancilla pair, and then explored the setup with all sites of the main chain coupled to ancillas.

Analysis of the density and entanglement dynamics in the single- and double-ancilla setups (Sec.~\ref{sec:repmeas}) reveals a ``quantum-Zeno-valve effect'' for sufficiently strong coupling $M$ between the ancilla and a site of the chain. This effects leads to long-lived quantum trajectories with very low entanglement corresponding to the blocking of the dynamics by a formation of a nearly bound state. This phenomenon is a manifestation of the interplay between the measurement backaction and the accumulated feedback of the measured system exerted on detectors in the absence of their reinitialization. 

As discussed in Secs.~\ref{sec:repmeas} and \ref{sec:5D}, the single- and double-ancilla setups already feature certain signatures of the many-ancilla entanglement transition. 
Specifically, we observe a crossover with increasing $M$ between two distinct types of behavior. For relatively small $M$, represented by $M=1$, all the quantum trajectories correspond to fast thermalization, with entanglement entropy approaching its maximum value and weak fluctuations. On the other hand, for large $M$, we have observed a strong variance of entanglement entropy, with some trajectories exhibiting strong entanglement across the system and others having strongly suppressed entanglement in comparison to thermal, chaotic states. This feature persists even for strong coupling $M = 50$, implying that typical trajectories are not representative for the dynamics at large $M$. The crossover between these types of dynamics occurs around $M=4$. Applying only a single (or double) measurement is not sufficient to disentangle all trajectories; we always observe at least a few trajectories that are strongly entangled, even for a very large measurement strength. Therefore, we conclude that a genuine entanglement transition requires a finite density of ancilla pairs. 

We find clear signatures of the transition in a setup where every site of the chain is measured, Sec.~\ref{sec:5D}. We have observed entangling behavior for weak measurement strength and disentangling behavior for strong measurement strength. 
Using matrix product states, we have studied the system sizes of up to $L=24$ sites in the main chain (ergo, $24 \times 3 = 72$ sites in the system as a whole). Analyzing the dependence of the entanglement entropy on time $t$ and the system size $L$, we have found evidence of a transition for a measurement strength $M \approx 4$. Remarkably, this is essentially the same value as was identified for the crossover in the setups with one or two ancillas. Our numerics provides indications of a volume-law entropy scaling in the entangling phase, which is in line with previous results on correlated chains, where different
implementations of measurements were employed~\cite{Fuji2020a,Goto2020a,Doggen2022a}. Note that such a behavior is also typical for the entanglement transition in random quantum circuits~\cite{Skinner2019a, Li2018a, Fisher2022}. 
Of course, a numerical study cannot rigorously exclude a possibility that the volume law holds only up to a certain large length scale where the entropy saturates (i.e., that there is no phase transition in the thermodynamic limit). We expect that future work (analytical and numerical) in this direction may help to address this uncertainty.

Our work demonstrates the capability of the developed approach to capture the physics of measurement-induced entanglement transitions. It sheds more light on the nature of the entanglement transition, by disclosing the tomographic signatures of the transition in terms of quantum trajectories.
Moreover, it reveals a high degree of  universality of the measurement-induced entanglement transition in correlated many-body systems with respect to implementation of measurements.

To conclude, let us list possible future directions opened by this work. 
By considering an intermediate probability $0 < P < 1$ of measuring the ancilla within a given measurement interval, 
or a reduced density of ancilla sites,
one can enrich the  phase diagram of a measurement-induced transition. Furthermore, a quantitative analysis of the measurement-induced transition and associated critical behavior in our framework remains to be performed. In addition, it would be interesting to explore how the resetting of ancillas affects the entanglement transition. It is also important to better understand the degree of universality of our findings, in particular, by exploring models with another form of coupling between the main system and the ancillas. Another possible avenue for future work is that in the noninteracting case $U = 0$, the volume-law behavior was argued to be unstable in the archetype measurement protocols. It is interesting to study whether this statement holds for our measurement setup, where the interaction between ancillas and the chain site is inevitably present even for $U=0$.  
Finally, the developed framework is expected to be valuable for applications in the context of quantum engineering and information processing, such as steering of quantum states. The present analysis may be adapted to many realistic interacting systems of interest, ranging from cold-atom or trapped-ion quantum simulators to superconducting qubit arrays. 

\section*{Acknowledgments}

We thank J.~Behrends, C.~Castelnovo, H.~Perrin, I.~Poboiko, P.~P\"opperl, and A.~\v{S}trkalj for useful discussions. 
EVHD and IVG gratefully acknowledge collaboration with Oleg Yevtushenko at early stages of this work. 
The work was supported by the Deutsche
Forschungsgemeinschaft (DFG): Project No. 277101999
-- TRR 183 (Project C01) and Grants No. EG 96/13-1 and No. GO 1405/6-1, by the Helmholtz International Fellow
    Award, and by the Israel Binational Science Foundation
    -- National Science Foundation through award
    DMR-2037654.
The simulations in this work have been performed using the \texttt{TeNPy} library, versions \texttt{0.7.2} and \texttt{0.8.4} \cite{tenpy}. We acknowledge support by the State of Baden-Württemberg through bwHPC.

\emph{Data availability.---} Numerical data is available from the corresponding author upon reasonable request. The source code can be obtained from \cite{DoggenProjAncCode}.

\appendix

\section{Numerical method} 
\label{sec:appendix_num}

For our simulations, we apply the procedure similar to the one outlined in Ref.~\cite{Doggen2022a}. That is, we use the time-dependent variational principle (TDVP) as applied to matrix product states (MPS) \cite{Haegeman2016a} to exactly compute the unitary evolution. The difference between the procedure outlined in Ref.~\cite{Doggen2022a} and the method used in this work, is that projections of the ancilla site are instantaneous. This is effected through the application of a strong imaginary on-site potential on the first ancilla site:
\begin{equation}
\label{Hproj}
    \mathcal{H}_\mathrm{proj} = \pm iA a_1^\dagger a_1.
\end{equation}

The dynamics is then computed over the time interval $[0,\eta]$, resulting in imaginary-time propagation. In the limit $A,\eta \rightarrow \infty$, this results in a projection onto either the $|01\rangle$ or $|10\rangle$ state for the ancilla, depending on the sign $\pm$ in Eq.~\eqref{Hproj}. As explained in the main text, this sign is determined stochastically, by applying the Born rule to the ancilla density $n_a$. During this process, the original Hamiltonian is ``switched off.'' In the case where there is more than one ancilla pair, each projection is performed consecutively, starting from the left side of the main chain ($i = 1$).

We choose the numerical parameters $A = 3 \cdot 10^8$ and $\eta = 10^{-6}$. Numerically, the procedure is subdivided into 10 smaller steps to improve stability. A larger value of $\eta$ would provide a closer approximation of the projection (typically, we find a deviation of around $10^{-4}$ in the ancilla density after projection); we choose a smaller value to reduce computational effort without significantly compromising accuracy.

For implementing the model using MPS, we map the system of the main chain plus the ancilla pairs onto a next-nearest-neighbor Hamiltonian, by ``folding'' the ancillary sites into the main chain. One can avoid introducing next-next-nearest neighbor terms by alternating the ancilla and main sites in the resulting one-dimensional chain---this structure facilitates the calculation. Note, however, that, because of this choice, it is not straightforward to find, e.g., multipartite entanglement involving the ancilla pair as one of the partitions. In the case of many ancillas (Sec.~\ref{sec:5D}) there is not sufficient ``space'' to fold both ancilla sites into the main chain, and we do use a next-next-nearest neighbor Hamiltonian.

We observe that, for the cases of one and two ancillas studied in Sec.~\ref{sec:repmeas}, there is no parameter range for which entanglement is strongly suppressed in all trajectories. Hence, there will be some trajectories poorly approximated in case the MPS is truncated. For that reason, we do not restrict the bond dimension of the MPS when dealing with one or two ancilla pairs; our approach then becomes equivalent to exact diagonalization and we are restricted to modest system sizes of $\approx 20$ sites. Alternatively, one could post-select only those trajectories that have limited entanglement, keeping the maximum bond dimension, at the cost of losing information about the more entangled trajectories. This might be an interesting avenue for future work.

 For the case of ancillas coupled to every site (Sec.~\ref{sec:5D}), the total number of sites in the whole system is up to $24 \times 3 = 72$. Obviously, an exact diagonalization is not possible any more.  Our analysis of the entanglement transition proceeds similarly to Ref.~\cite{Doggen2022a}. We perform the calculations with the bond dimensions $\chi=128$ and 256. For sufficiently strong coupling to ancillas, $M \ge 4$, \emph{all} quantum trajectories exhibit suppressed entanglement, so that the truncation of the bond dimension to the lower-entanglement subspace does not influence the result. On the other hand, for weaker couplings, $M \le 3$, the bond-dimension truncation starts limiting the growth of the entanglement at long times, as pointed out in the main text.

\begin{figure}
    \centering
    \includegraphics[width=\columnwidth]{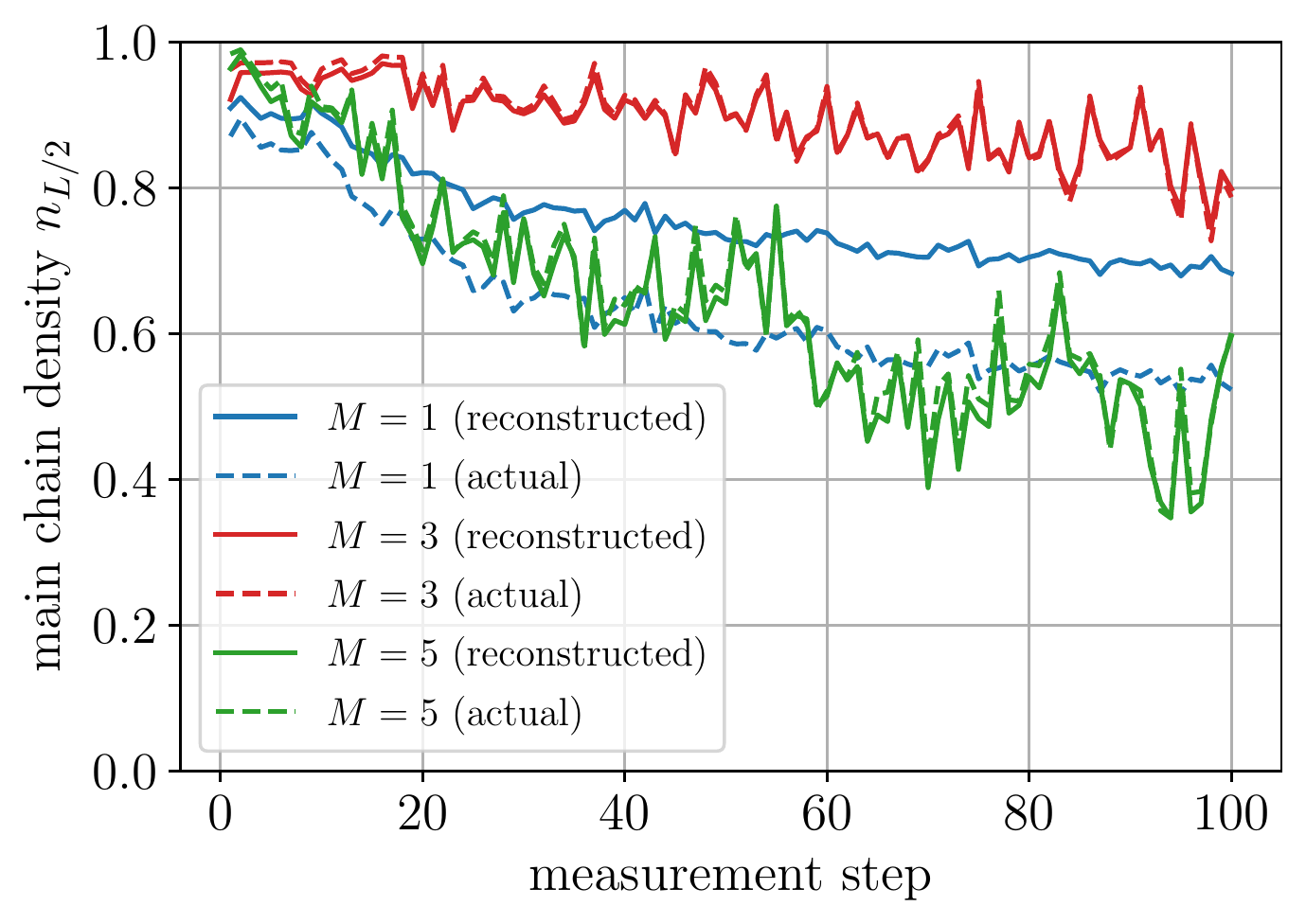}
    \caption{Reconstruction of the local density in the main chain, $n_{L/2}$, as inferred from Rabi oscillations in the ancilla attached to the measured site of the chain, for various choices of the measurement strength $M = 1, 3, 5$. For each measurement step of duration $\Delta T = 2$, the reconstructed (solid curves) density is compared to the actual density (dashed curves) obtained from the numerical simulation of the many-body dynamics. Note that $n_{L/2}$ deviates from its initial value $n_{L/2}(t=0)=1$ rather slowly for $M=3$, as compared to $M=1$ and $M=5$. This is a manifestation of a partial freezing induced by an effectively strong measurement near the Rabi resonance: the quantum-Zeno-valve effect. For $M = 3$ and $M = 5$ the curves are almost on top of each other for most measurement steps. }
    \label{fig:Rabitime}
\end{figure}

\begin{figure*}
    \centering
    \includegraphics[width=.9\textwidth]{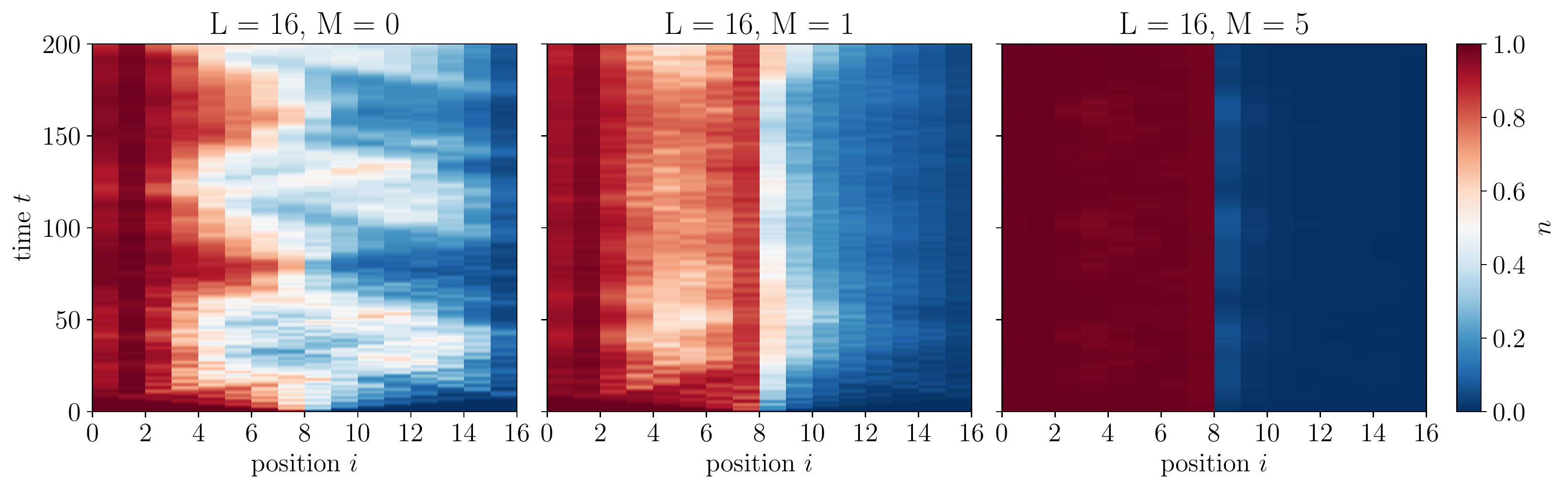}
    \caption{Dynamics of the particle density $n$ in the main chain for the single-ancilla case, starting from the domain-wall initial state (Fig.~\ref{fig:diag}), where the time evolution is computed without any projective measurements $\Delta T \rightarrow \infty$, for the uncoupled case $M = 0$ (left panel), $M = 1$ (middle panel), and $M = 5$ (right panel).}
    \label{fig:M0}
\end{figure*}

\section{Protocol for the reconstruction of the density in the main chain}
\label{sec:appendix_meas}

In the main text, Sec.~\ref{sec:rabi}, we have discussed the effect of the main chain on the ancilla dynamics. As a ``proof of concept," let us provide a demonstration of how one could infer the main chain density $n_i$ from the ancilla dynamics coupled to the site $i$. For the same parameters as used in Fig.~\ref{fig:Rabi}, we consider 100 measurement cycles, i.e., continue the runs up to time $t = 200$. For each cycle, we infer the density on the chain site connected to the ancilla by fitting Eqs.~\eqref{eq:rabi} and \eqref{eq:rabiamp} to the ancilla density. Here, we only consider those cases where a particle was measured in the ancilla, separately for each measurement step. For sufficiently coupling strength $M$, this procedure provides remarkably good agreement to the true density (as numerically obtained from the full many-body wave function). For $M = 3$ and $M = 5$, the two curves match very closely, while for weaker measurement strength $M = 1$, in which case the two-level approximation does not work as well, the agreement is qualitative. Thus, the employed ancilla-based setup is indeed capable of measuring the true density in the chain.

Of course, in a realistic experimental setting, we do not have direct access to the ancilla dynamics $n_a(t)$ while measurements are not performed, as such measurements would necessarily be at least partially destructive and affect the ancilla dynamics itself. However, Fig.~\ref{fig:Rabitime} does clearly demonstrate that the information about the main chain density $n_{L/2}$ is strongly correlated to the ancilla density. Thus, it is in principle possible to obtain this information from projective measurements. One possible approach is to restrict the measurement interval $\Delta T$ in such a way that it is smaller than half a period of the ancilla Rabi oscillations for the non-measured case \mbox{$M = 0$}, leading to the simple criterion $J\Delta T < \pi$. Then, a one-to-one mapping of $n_a$ to $n_{L/2}$ can be constructed by cross-correlating the results of projective measurements for many different choices of $M$ over several trajectories.

\begin{figure*}[!htb]
    \centering
    \includegraphics[width=.31\textwidth]{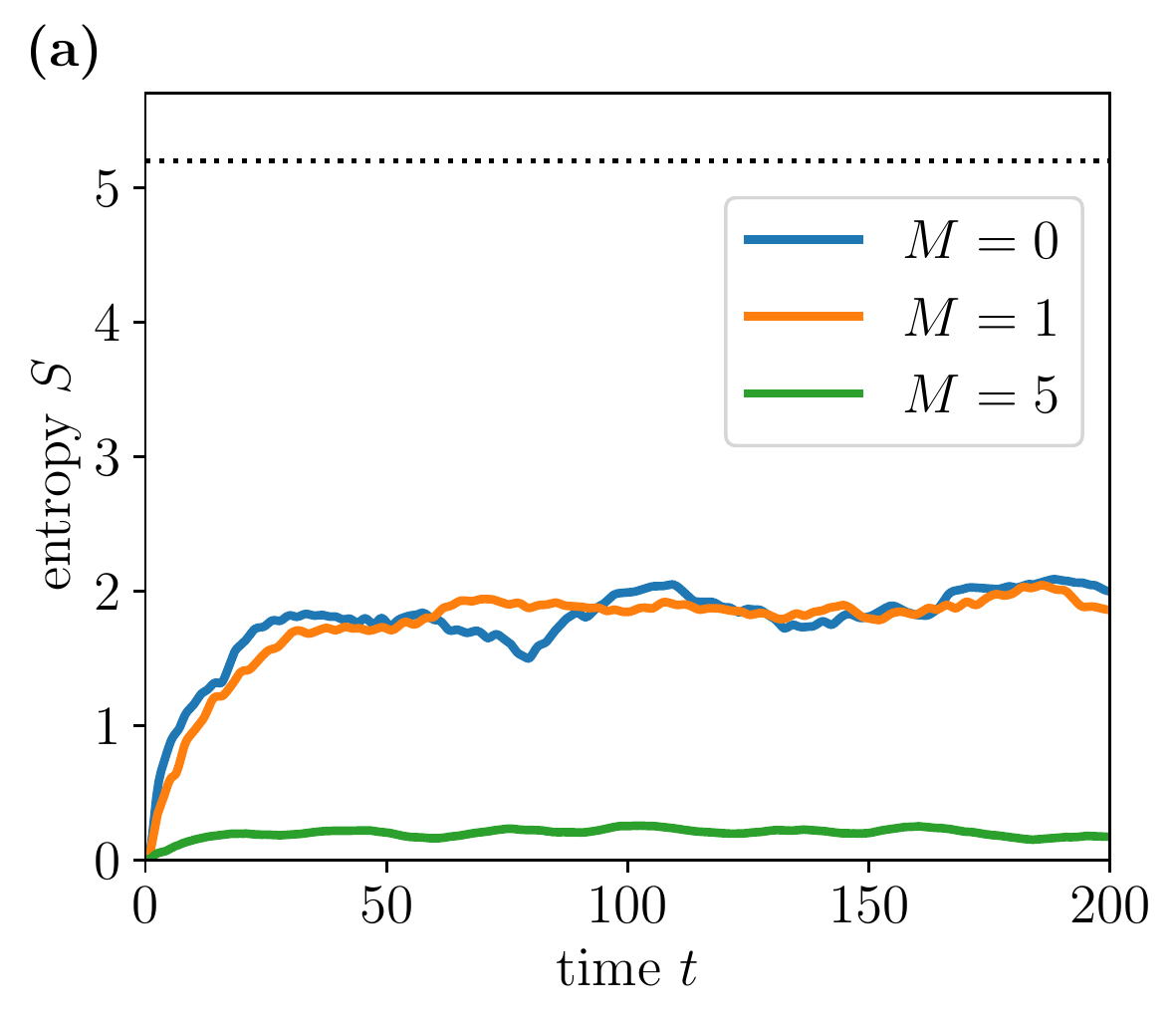}
    \includegraphics[width=.33\textwidth]{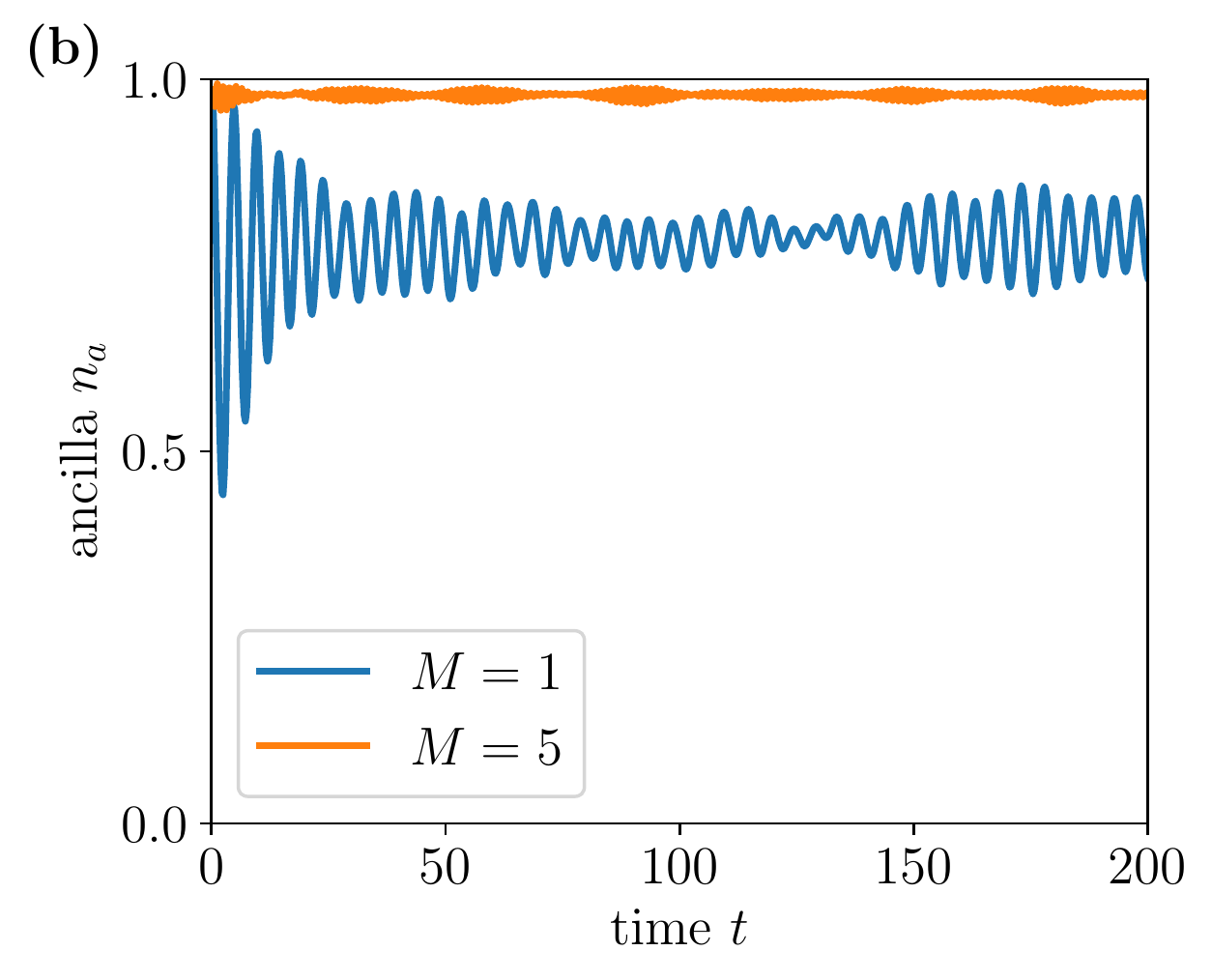}
    \includegraphics[width=.31\textwidth]{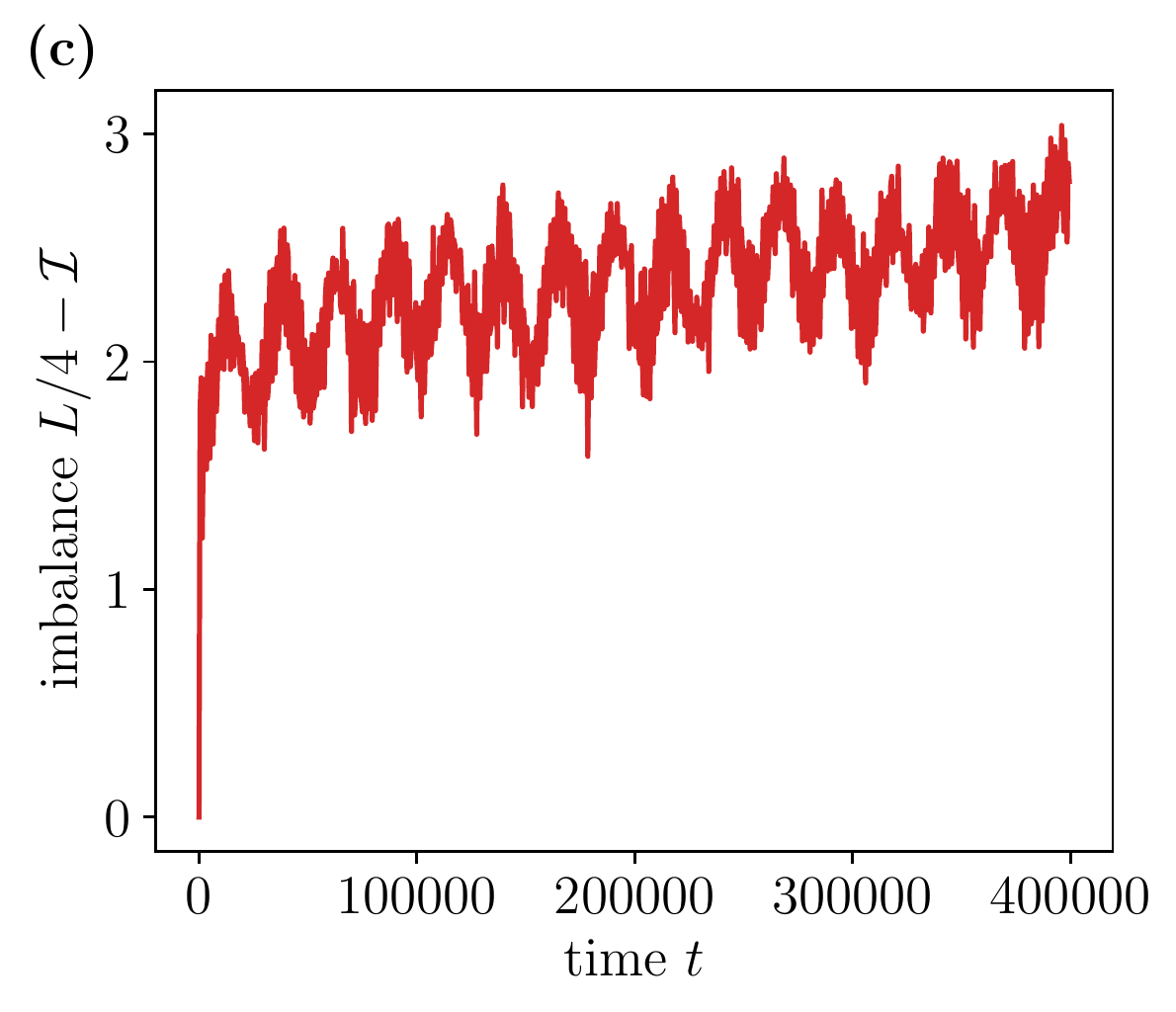}
    \caption{\textbf{(a)}: The von Neumann entropy of entanglement $S$ as a function of time for the case without projective measurements,  $\Delta T \rightarrow \infty$, using the single-ancilla setup (Fig.~\ref{fig:diag}) with a domain-wall initial state, for the uncoupled case $M = 0$ and the coupled cases $M = 1$ and $M=5$. The system length is $L=16$, as in  Fig.~\ref{fig:M0}. The dotted line indicates the Page entropy corresponding to a thermal state. \textbf{(b)}: Dynamics of the ancilla density $n_a$ for the non-measured case $\Delta T \rightarrow \infty$, starting from the domain-wall initial state in the single-ancilla setup (Fig.~\ref{fig:diag}) with $L=16$. \textbf{(c)}: 
    Dynamics for the case $L = 14$ and $M=0$, up to a very long time $t = 4 \cdot 10^5$. Shown is the imbalance in the main chain, which slowly trends toward the thermal value $3.5$.
    }
    \label{fig:app_nomeas}
\end{figure*}

\section{Dynamics in the absence of measurements}

In the main text, we have discussed the case where the main chain of the system is coupled to one or more ancilla pairs with a nearest-neighbor interaction $M$, and projective measurements are ``stroboscopically'' performed at regular intervals $\Delta T$. For the sake of comparison, it is useful to consider the cases where there is no interaction, and where there is coupling to ancilla but no measurement is performed on the ancilla. We will consider both cases in this Appendix.

\subsection{Uncoupled case} \label{sec:appendix_unc}

Let us first compare the dynamics of the measured system to that in the ``uncoupled'' case without any interaction between the main chain and the ancilla, $M = 0$. We show the result in the left panel of Fig.~\ref{fig:M0}, starting from the domain-wall initial state. In this case, the density evolution is believed to feature integrable Kardar-Parisi-Zhang dynamics \cite{ljubotina17,Misguich2017a,ljubotina19,Wei2022a}. Interestingly, in the finite-size setup, we observe multi-scale oscillations of the density, apparently related to the interference effects resulting from multiple bounces of the excitations from the boundaries in this integrable model. The analytical description of this phenomenon is, however, beyond the scope of this work. Comparing to the case of repeated measurements as shown in the main text (Sec.~\ref{sec:repmeas}), we see that a finite ancilla-chain coupling $M$ in a single-ancilla setup destroys the peculiar features of Kardar-Parisi-Zhang (KPZ) domain-wall melting by breaking the integrability of the system.

\begin{figure*}
 \centering
 \includegraphics[width=.9\textwidth]{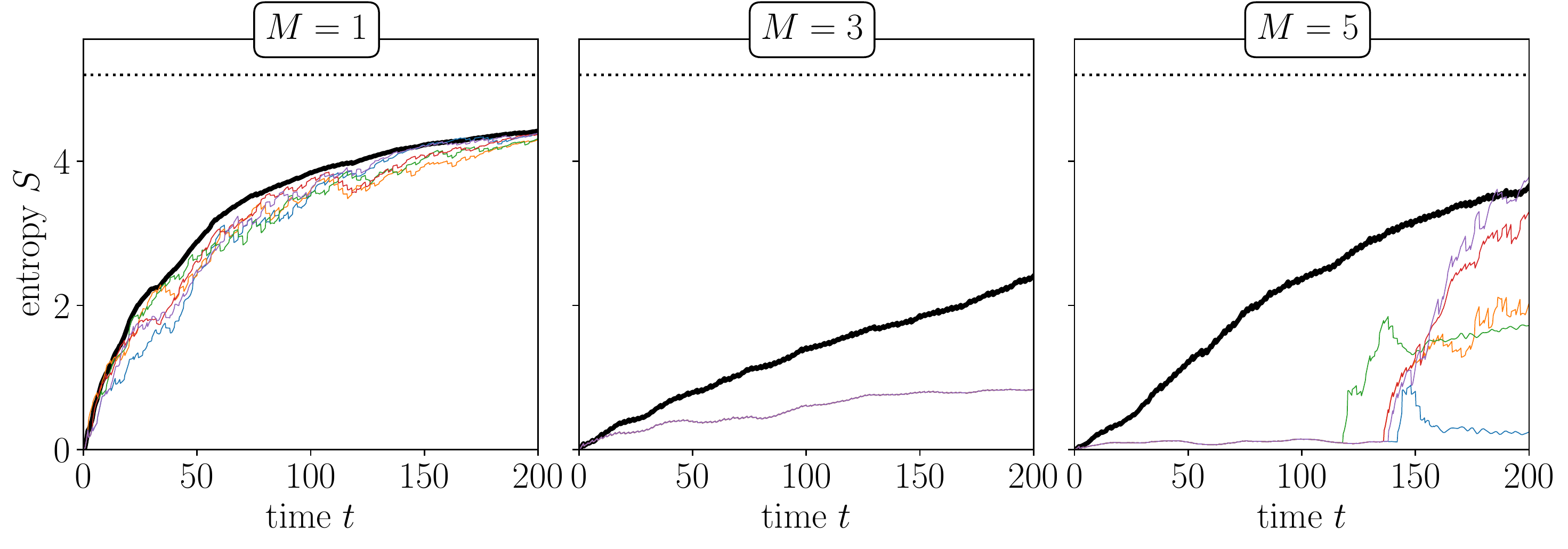}
 \caption{Entropy dynamics for the domain-wall initial condition, as in Fig.~\ref{fig:ms8L16}(c) of the main text. By contrast, here we show the five curves (colored lines) with the \emph{smallest} integrated entropy $\Xi$, alongside the average (thick black line). The number of trajectories $R = 40$.}
 \label{fig:leastentangled}
\end{figure*}

To provide a further benchmark, we have compared the integrable dynamics computed using our TDVP implementation at $M = 0$ (decoupled ancilla) to the dynamics obtained from the exact simulation of the main-chain dynamics only (i.e., no ancilla present) as obtained from a fully independent calculation using the \texttt{QuSpin} library \cite{QuSpin2019}, finding excellent agreement. The result for the imbalance dynamics up to a very long time $t = 400000$ is shown in Fig.~\ref{fig:app_nomeas}(c). Here the KPZ domain-wall melting is the steep initial increase. It is followed by a slow creep and oscillations, which are determined by boundary effects. This long-time behavior remains to be explained; this is however beyond the scope of the present paper.

\subsection{No ancilla projections}
\label{aC}

It is also useful to investigate the case where the system is quenched at $t = 0$, but no measurement is performed. Let us  consider the case of a single ancilla with the domain-wall initial condition. The results for $M = 1$ and $M = 5$ are shown in the middle and right panels of Fig.~\ref{fig:M0}, the corresponding entropy dynamics is shown in Fig.~\ref{fig:app_nomeas}(a), and the dynamics of the ancilla is shown in Fig.~\ref{fig:app_nomeas}(b). Comparing the case of only this single quench to the case of repeated measurements (Fig.~\ref{fig:ms8L16} of the main text), we see dramatic differences. Indeed, in the absence of measurements, the state remains close to its initial  form: the density distribution remains close to the domain-wall initial condition, the entropy stays well below the Page value, and the ancilla density $n_a$ stays close to unity (well above $1/2$). The mechanism of this freezing is the formation of a repulsively bound state, as discussed in the main text. On the other hand, repeated measurements lead to thermalization that manifests itself in the entropy approaching the Page value, domain wall smearing, and the ancilla density relaxing towards $1/2$ (cf. Fig.~\ref{fig:ms8L16}). This demonstrates that the physics studied in this paper is not just governed by the coupling to an ancilla but is rather induced by projective measurements of the ancilla degrees of freedom.

\section{Additional data for the single-ancilla setup with a domain-wall initial state}
\label{sec:appendix_dwall}

In Sec.~\ref{sec:singlepair}, we have studied the dynamics of the entanglement entropy and its fluctuations for the setup with one ancilla and a domain-wall initial state. In this Appendix, we provide additional numerical data for this setup. 

In Fig.~\ref{fig:leastentangled}, the five least entangled trajectories, as quantified by the integrated entropy $\Xi$ [Eq.~\eqref{eq:entang_sorting}], are displayed. 
For the case of relatively weak measurement strength, $M = 1$, all curves are similar, with only minor fluctuations, and the least entangled trajectories are close to the average, as is also expected from the corresponding data in Fig.~\ref{fig:ms8L16}(c). For the resonant case $M = 3$, we see that all curves fall on top of each other, implying that the measurement results are identical for each of these trajectories. For a still stronger (and non-resonant) coupling, $M = 5$, all five trajectories are identical (and are characterized by a very small entropy) up to a time $t \approx 120$, after which they start exhibiting branching off towards high entropy state. These three panels provide an additional manifestation of a distinct physics in the case of relatively weak ($M=1$), resonant ($M=3$), and strong off-resonant ($M=5$) couplings.

\begin{figure}
    \centering    \includegraphics[width=\columnwidth]{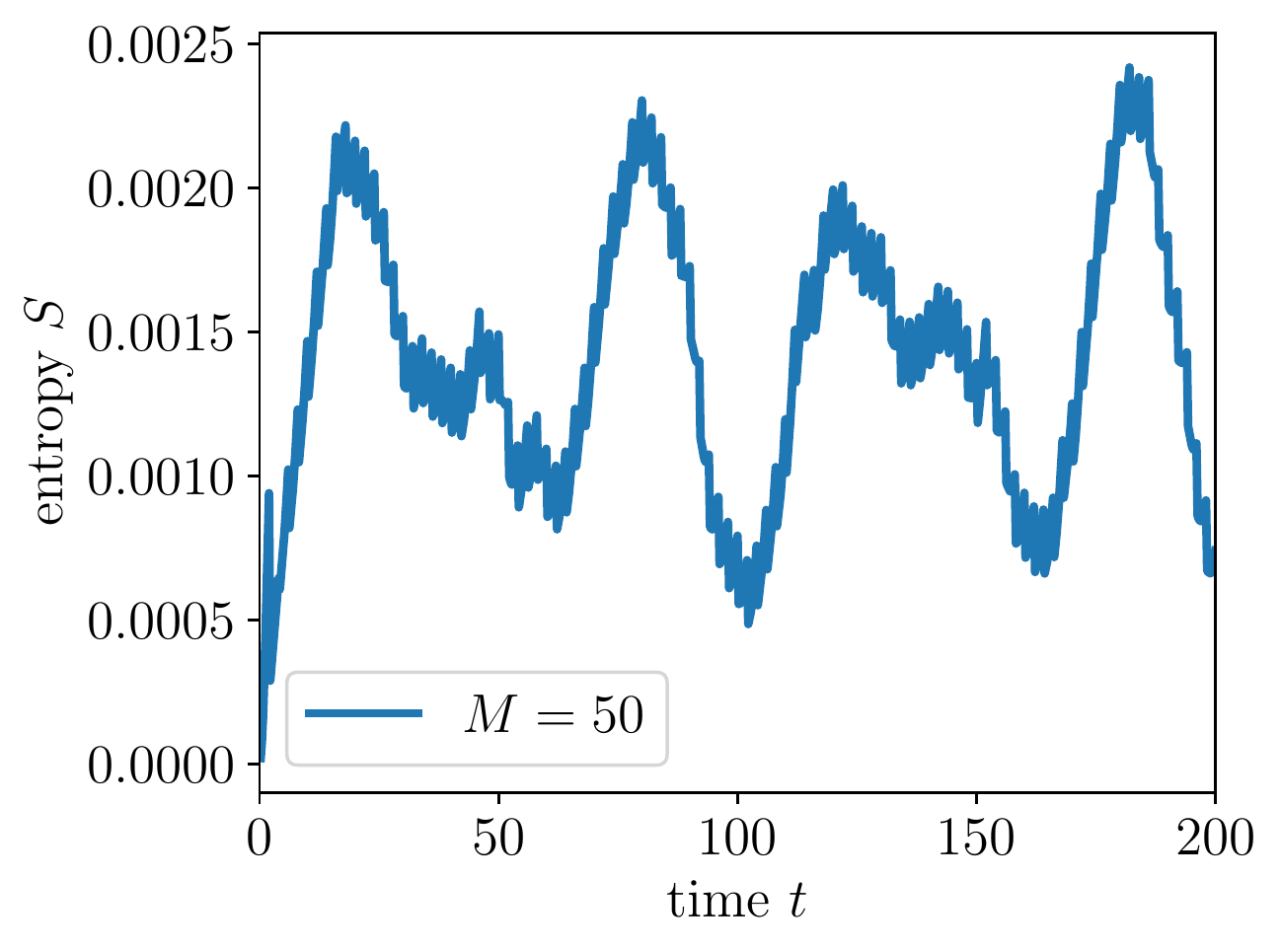}
    \caption{Dynamics of the von Neumann entropy $S$ as a function of time and averaged over $R = 40$ trajectories. The measurement strength $M = 50$ and we consider the domain-wall setup with a single ancilla, as in Fig.~\ref{fig:ms8L16}c of the main text. Note the scale on the $y$-axis: the values of $S$ are very small.}
    \label{fig:domainwall_M50}
\end{figure}

We also show in Fig.~\ref{fig:domainwall_M50} the entropy dynamics for the case $M = 50$ in the domain-wall setup. The data complement those shown in Fig.~\ref{fig:ms8L16}c for the same setup and $M= 1$, 3, and 5.  As pointed out in the main text, the dynamics here almost totally freezes for \emph{all} trajectories since the Rabi oscillation amplitude approaches zero as $M \rightarrow \infty$ and a ``nearly bound state'' is formed. The domain wall is therefore robust up to very long times. This manifests itself in very low values of the entanglement entropy for all trajectories.

\bibliography{ref}

\end{document}